\documentclass[12pt]{iopart}
\usepackage{epsfig,harvard}

\begin{document}
\title[e-H ionization]{Electron-impact ionization of atomic hydrogen from near
threshold to high energies}
\author{Igor Bray
\thanks{electronic address: I.Bray@flinders.edu.au}
}
\address{
Electronic Structure of Materials Centre,
The Flinders University of South Australia,
G.P.O. Box 2100, Adelaide 5001, Australia}
\date{\today}

\begin{abstract}
Application of the convergent close-coupling (CCC) method to
electron-impact ionization of the ground state of atomic hydrogen is
considered at incident energies of 15.6, 17.6, 20, 25, 27.2, 30, 54.4,
150 and 250~eV. Total through to fully differential cross sections are
presented. Following the analysis of Stelbovics [submitted to
Phys.~Rev.~Lett. (http://xxx.lanl.gov/abs/physics/9905020)] the
equal-energy sharing cross sections are calculated using a solely
coherent combination of total-spin-dependent ionization amplitudes,
which are found to be simply a factor of two greater than the
incoherent combination suggested by Bray and Fursa [1996 {\em
Phys.~Rev.~A} {\bf 54},~2991]. As a consequence, the CCC theory is
particularly suited to the equal-energy-sharing kinematical region,
and is able to obtain convergent absolute scattering amplitudes, fully
ab initio. This is consistent with the step-function hypothesis of
Bray [1997 {\em Phys.~Rev.~Lett.} {\bf 78},~4721], and indicates that
at equal-energy-sharing the CCC amplitudes converge to half the step
size.  Comparison with experiment is satisfactory in some cases and
substantial discrepancies are identified in others. The discrepancies
are generally unpredictable and some internal inconsistencies in the
experimental data are identified.  Accordingly, new (e,2e)
measurements are requested.
\end{abstract}
\jl{2}
\submitted
\maketitle

\section{Introduction}
Our primary motivation in the study of electron-atom interactions is
to provide accurate data for the needs of science and industry. To
this end the primary emphasis of our study has been on discrete
excitation processes. The locally developed convergent close-coupling
(CCC) method was aimed at resolving the long-standing discrepancy of
the elementary electron-impact 2P excitation in atomic
hydrogen~\cite{BS92}. The basic idea was the same as used earlier in,
for example,
pseudostate close-coupling method of \citeasnoun{WW86}, except that
the generation of the pseudostates was done using an orthogonal
Laguerre basis. This allowed for a systematic study of convergence in
the observable of interest (eg. 2P excitation) with increasing number
of states $N$. The association of the pseudostates with an equivalent
quadrature 
rule for the infinite sum and integral over the true target discrete
and continuum spectrum indicated the importance of an efficient numerical
implementation that allowed for coupling of as many states as possible 
for given computational resources.

Our interest in ionization has come about rather indirectly. First we
noted that the CCC method was able to reproduce the total e-H
ionization cross section~\cite{BS93l}. This cross section was obtained 
by essentially summing the cross sections for excitation of the
positive-energy pseudostates, thereby identifying excitation of these
states with ionization processes. Another important indication that
the CCC method, and close-coupling approaches generally, should be
able to obtain accurate ionization cross sections was found by
application to 3P excitation of sodium~\cite{B94R}. It was found that
in order to be certain of obtaining accurate scattering amplitudes, for 
even the sodium 3P excitation, coupling within the ionization channels had to 
be treated accurately. This was a most unexpected result with the
consequence of our direct interest in ionization processes.

In recent years considerable progress has been made in the ability of
theory to reproduce fully differential measurements  of atomic
electron-impact ionization. There are a number of theoretical
approaches. Some approach the problem from the asymptotically
correct three-body boundary conditions
\cite{BBK89,BBB91,BB94,B97,Chen_etal98,JM98}. Others are based on the
Born approximation with the introduction of distorting and other
potentials to improve the accuracy at lower energies
\cite{PS91,PS92,JMS92,WAW93,Whelan94,Roeder96}. More recently, a new
and very promising development involves evaluation of ionization
without reference to asymptotic boundary conditions
\cite{MRB97,BRIM99}. There are also time-dependent approaches
\cite{IDHF95,PS96}. 
Another approach attempts to solve the Schr\"odinger equation of the
scattering system subject to the approximation that the total wave
function is expanded in a finite set of square-integrable target states 
\cite{CW87,CWW91,BKMS94}. It is the latter approach that is of
particular interest to us. It allows for the treatment 
of discrete excitation and ionization simultaneously, which to our
mind is necessary to be sure of the accuracy of either calculation.

The CCC theory has been extensively applied to e-He ionization at high 
\cite{BF96} intermediate \cite{REBFM96,REBFM96l,BFRE97_64,Rioual98} and low
\cite{BFRE98} energies. Most encouraging was the ability to accurately 
describe both excitation and ionization 100~eV data using a single CCC
calculation \cite{BF96l}. During the course of this study some
difficulties relating to the accuracy of absolute differential
ionization cross sections were identified and studied systematically
\cite{REBF97}. It was determined that with decreasing projectile
energy the singly differential cross section (SDCS) develops unphysical
oscillations, which in turn affect the magnitude of the
angle-differential ionization cross sections, though apparently not
their angular distributions. The source of this problem was suggested
to be due the fact 
that for infinite $N$ the CCC-calculated SDCS at any total (excess)
energy $E$ should only yield
physically meaningful results on $[0,E/2]$ secondary energy range
and zero elsewhere \cite{B97l}. In other words, with increasing $N$
the CCC-calculated SDCS should converge to a step-function. Though
this is a conceptually useful result, as it allows unambiguous
identification of the physical scattering amplitudes, in practice for
small-enough $E$,
finite calculations yield oscillatory SDCS and there is small but
nonzero flux at secondary energies greater than $E/2$. It is our
view that  this is a fundamental limitation of the close-coupling approach to
ionization. 

Nevertheless, the utility of the approach at low energies is not as
diminished as one might at first suspect. The reason why the angular
distributions were relatively unaffected, except by an overall factor
obtained from the SDCS, was related to the equivalent-quadrature idea
of the pseudostates and an empirical scheme for
choosing the states was given \cite{B99jpbl}. Consequently, if the
true SDCS was available then rescaling all of the angle-differential
ionization cross sections by the ratio of the true to the
CCC-calculated SDCS would result in relatively accurate magnitudes
also. This idea has been applied successfully to helium, where
rescaling factors of approximately two were identified and brought
about good agreement with experiment \cite{BFRE98,Rioual98}. Not so in 
the case of ionization of atomic hydrogen by 15.6~eV electrons
\cite{B99jpbl}, where the estimated rescaling by 2.7 still left the theory a
factor of two or so less than experiment.

To complicate things further \citeasnoun{BC99} have questioned the
validity of the CCC approach to ionization at any energy. They claim
that the CCC-calculated ionization scattering amplitudes as defined by 
\citeasnoun{BF96} should satisfy the symmetrization postulate
\begin{equation}
f_S(\bi{k},\bi{q})=(-1)^Sf_S(\bi{q},\bi{k})
\label{sympos}
\end{equation}
for e-H ionization, where $S$ is the total spin. The fact that 
they don't (CCC-calculated SDCS is not symmetric about $E/2$) they take
to indicate a lack of convergence everywhere, 
and presumably, agreement with experiment is coincidental. 

Another
criticism of our work relates to the incoherent combination of
CCC-calculated amplitudes on either side of $E/2$. Whereas this choice 
was taken in order to retain the unitarity of the close-coupling
formalism \citeasnoun{S99l} showed that this was not necessary. By
studying the S-wave model he showed that the CCC-calculated ionization
amplitude was able to be clearly defined only for $k=q$ and
the cross section should be given by
\begin{equation}
\frac{d^3\sigma_S}{d\Omega_1d\Omega_2dE_2}=
|f_S^{(N)}(\bi{k},\bi{q})+(-1)^Sf_S^{(N)}(\bi{q},\bi{k})|^2,
\label{XSECA}
\end{equation}
as opposed to the prescription given by \citeasnoun{BF96}
\begin{equation}
\frac{d^3\sigma_S}{d\Omega_1d\Omega_2dE_2}=
|f_S^{(N)}(\bi{k},\bi{q})|^2+|f_S^{(N)}(\bi{q},\bi{k})|^2,
\label{XSECB}
\end{equation}
where the $f_S^{(N)}$ are the amplitudes calculated in the CCC
theory. \citeasnoun{S99l} also concluded that apparent  convergence of
the CCC results at $E/2$ was real and that it was to half the true
scattering amplitude, or one quarter the true cross section. The
consequence of his work is profound. It suggests that the CCC method
is ideal for equal-energy-sharing kinematics where it is able to yield
convergent cross sections in both shape and magnitude fully ab initio
without any reference to rescaling.

To address these issues we perform a systematic study of e-H
ionization from high energies through to low. We give our best
estimates for the total through to fully differential ionization cross
sections and discuss the issues involved.

\section{Theory}
The details of the CCC theory have already been given
\cite{BF96}. Here we outline some of the major issues of interest.
We begin with the standard Born approximation because it is accurate
at high energies and the objections raised by \citeasnoun{BC99} are
equally applicable to our interpretation of this approximation.
Unless specified otherwise atomic units are assumed throughout.
\subsection{The Born approximation}
If one needs a quick approximate estimate of an excitation scattering
process then the Born approximation is an excellent candidate as it
covers an immense energy range. The total Hamiltonian $H$
is partitioned asymmetrically
$H=K+V$, where $K=K_1+H_2$ is the asymptotic Hamiltonian, and where
$K_1$ is the free projectile kinetic energy operator and $H_2=K_2+V_2$
is the hydrogen target Hamiltonian. The projectile(target) potential
is $V_1$($V_2$), and $V=V_1+V_{12}$ is the asymptotic potential, where
$V_{12}$ is the electron-electron potential.

The differential cross section for excitation of the hydrogen
ground state $\phi_i$ to state $\phi_f$ by an electron of incident
momentum $\bi{k}_i$ is approximated via
\begin{equation}
\frac{d\sigma_{fi}}{d\Omega}\approx
|\langle \bi{k}_f(1)\phi_f(2)|V|\phi_i(2)\bi{k}_i(1)\rangle|^2,
\label{bornd}
\end{equation}
where the channel states satisfy
\begin{equation}
K|\phi_n(2)\bi{k}_n(1)\rangle=(\epsilon_n+k^2_n/2)|\phi_n(2)\bi{
k}_n(1)\rangle.
\end{equation}
In the Born
approximation the total wavefunction is simply written as 
\begin{equation}
|\Psi_{Si}^{(+)}\rangle\approx|\phi_i\bi{k}_i\rangle,
\label{bornpsi}
\end{equation}
which neglects antisymmetry (has no dependence on total spin $S$) or
coupling to other channels.

The Born
approximation may also be readily applied to ionizing collisions,
for total energy $E=\epsilon_i+k_i^2/2>0$, by
simply replacing the discrete eigenstate $\phi_f$ in \eref{bornd} with
a continuum eigenstate $\bi{q}^{(-)}_f$, a Coulomb wave of momentum 
$\bi{q}_f$ and energy $q_f^2/2=E-k_f^2/2$. Then the triply (fully)
differential cross section (TDCS) may be written as
\begin{equation}
\frac{d^3\sigma}{d\Omega_1d\Omega_2dE_2}\approx
|\langle \bi{k}_f\bi{q}^{(-)}_f|V|\phi_i\bi{k}_i\rangle|^2,
\label{bornc1}
\end{equation}
Immediately we run into a problem. In the case of
ionization we have two electrons going out with momenta $\bi{k}_f$ and
$\bi{q}_f$, typically one much faster
than the other. Which one do we assign to the plane wave (electron
one) and which to 
the Coulomb wave (electron two)? 
Numerical investigation shows that the slower
electron should be treated as a Coulomb wave and the faster as a plane 
wave. This is often justified as a shielding approximation: the fast
electron is shielded from the proton by the slow electron, which in
turn moves in the potential of the proton. While this
seems very sensible we find it somewhat a mixture of ideas given that
the time-independent formalism is being used. Instead, we suggest that 
a more consistent approach is to state that there are two
theoretically distinguishable Born approximations to any ionization
process which must be combined incoherently
\begin{equation}
\frac{d^3\sigma}{d\Omega_1d\Omega_2dE_2}\approx
|\langle \bi{k}_f\bi{q}^{(-)}_f|V|\phi_i\bi{
k}_i\rangle|^2+
|\langle \bi{q}_{f'}\bi{k}^{(-)}_{f'}|V|\phi_i\bi{k}_i\rangle|^2.
\label{bornc2}
\end{equation}
Here the plane wave $\langle\bi{q}_{f'}|$ has momentum $\bi{
q}_{f'}=\bi{q}_f$
and Coulomb wave $\langle\bi{k}_{f'}^{(-)}|$ has momentum
$\bi{k}_{f'}=\bi{k}_f$. We use the $f'$ index to ensure clear
distinguishability between the two cases.

Without loss of generality let us suppose that $q_f\le k_f$. 
For $q_f^2/2\ll k_f^2/2$ there is no difference between \eref{bornc1} and
\eref{bornc2} since the first term in \eref{bornc2} is typically an
order of magnitude or more bigger than the second. 
Though the Born approximation works well for such cases, we may wish to
apply it at low energies to say demonstrate the difference between
the Born approximation and a more realistic theory. In this case
the two terms may be of similar magnitude, and we feel
that \eref{bornc2} is a more consistent interpretation of the Born
approximation utilising solely the rules of non-stationary Quantum
Mechanics. 

The primary advantage of the formulation \eref{bornc2} is that of
clarity. For example, how should the total ionization cross section be 
defined following the definition \eref{bornc1}? Should the endpoint of 
the energy integration ($dE_2$) be $E/2$ or $E$? This question is
worthwhile addressing even if in practice the energy integral typically
converges well before $E/2$. From our perspective, for the Born
approximation \eref{bornc1} (no antisymmetry, electrons are
distinguishable) the 
endpoint of the integration to form the total
ionization cross section $\sigma_I$ should be $E$, i.e.
\begin{eqnarray}
\sigma_I&\approx&\int_0^E dE_2\int d\Omega_1 d\Omega_2
|\langle \bi{k}_f\bi{q}^{(-)}_f|V|\phi_i\bi{
k}_i\rangle|^2\nonumber\\\label{intE}
&\equiv&\int_0^{E} dE_2 \frac{d\sigma}{dE_2}(E_2)\\\label{intE/2}
&=&\int_0^{E/2} dE_2 \left[\frac{d\sigma}{dE_2}(E_2) +
\frac{d\sigma}{dE_2}(E-E_2)\right]\\ 
&\equiv&\int_0^{E/2} dE_2\int d\Omega_1 d\Omega_2 
\left(|\langle \bi{k}_f\bi{q}^{(-)}_f|V|\phi_i\bi{
k}_i\rangle|^2+
|\langle \bi{q}_{f'}\bi{k}^{(-)}_{f'}|V|\phi_i\bi{
k}_i\rangle|^2\right),
\label{TICS}
\end{eqnarray}
where $E_2=q_f^2/2$. Thus, as far as the Born approximation to
e-H ionization is concerned, we suggest that as the difference in the
energies of the two outgoing electrons increases the first term in
\eref{bornc2} and \eref{TICS} converges to the true scattering
amplitude, whereas the second converges to zero.

The objections of \citeasnoun{BC99} are applicable to our
interpretation of the Born approximation. The symmetrization postulate
\eref{sympos} 
is not satisfied as there is no spin-dependence. The two terms are
combined without any normalisation factors, since any such factor
would affect the first, most dominant term. Nevertheless, the Born
approximation has value over an immense kinematical range.

\subsection{The close-coupling with no exchange approximation}
To improve on the Born approximation we need to allow for coupling to
other channels and antisymmetry of the total wave function. We
consider the former first. Improvement on \eref{bornpsi} is provided by
the approximation
\begin{eqnarray}
|\Psi_{Si}^{(+)}\rangle\approx|\Psi_{i}^{(+)}\rangle&\approx&
I_2^{(N)}|\Psi_i^{(+)}\nonumber\\
&=&\sum_{n=1}^N|\phi_n^{(N)}\rangle\langle\phi_n^{(N)}
|\Psi_i^{(+)}\rangle\nonumber\\
&=&\sum_{n=1}^N|\phi_n^{(N)}f_{ni}^{(N+)}\rangle,
\end{eqnarray}
where the $N$ functions $\phi_n^{(N)}$ form an orthonormal set, and
the functions $f_{ni}^{(N+)}$ we obtain by solving the spin-independent
close-coupling equations for the $T$ matrix
\begin{eqnarray}
\langle\bi{k}_f\phi_f^{(N)}|T|\phi_i^{(N)}\bi{k}_i\rangle &\equiv&
\langle\bi{
k}_f\phi_f^{(N)}|V|\sum_{n=1}^N\phi_n^{(N)}f_{ni}^{(N+)}\rangle\nonumber\\ 
&=&
\langle\bi{k}_f\phi_f^{(N)}|V|\phi_i^{(N)}\bi{k}_i\rangle\nonumber\\ 
&&+\sum_{n=1}^N\int d^3k\frac{\langle\bi{
k}_f\phi_f^{(N)}|V|\phi_n^{(N)}\bi{k}\rangle 
\langle\bi{k}\phi_n^{(N)}|T|\phi_i^{(N)}\bi{k}_i\rangle}
{E+i0-\epsilon_n^{(N)}-k^2/2}.
\label{TLS}
\end{eqnarray}
The expansion states $\phi_n^{(N)}$ must be square-integrable in order 
that all of the $V$-matrix elements were calculable. Furthermore, we
desire that
\begin{equation}
\lim_{N\to\infty}I_2^{(N)}|\Psi_i^{(+)}\rangle=
I_2|\Psi_i^{(+)}\rangle=|\Psi_i^{(+)}\rangle,
\end{equation}
where $I_2$ is the true target-space identity operator. This may be
achieved by diagonalising the target Hamiltonian $H_2=K_2+V_2$
using a Laguerre basis to yield $\phi_n^{(N)}$ such that
\begin{equation}
\langle\phi_f^{(N)}|H_2|\phi_i^{(N)}\rangle=\delta_{fi}\epsilon_f^{(N)}.
\label{diag}
\end{equation}
The diagonalization \eref{diag} results in states with negative and
positive energies. With increasing $N$ the negative energy states
$\phi_f^{(N)}\to\phi_f$, the true discrete eigenstates, and the
positive energy states provide an increasingly dense discretization of 
the continuum.

The close-coupling approximation (without exchange) builds on top of the Born
approximation and so has the same asymptotic Hamiltonian and channel
functions. It is unitary and the sum over $n$ implies an on-shell
integration over the continuum from zero to total energy $E$.
The transition matrix is 
\begin{eqnarray}
\langle\bi{k}_f\phi_f|T|\phi_i\bi{k}_i\rangle\equiv
\langle\bi{k}_f\phi_f|V|\Psi_{Si}^{(+)}\rangle
&\approx&\langle\bi{k}_f\phi_f|I_2^{(N)}VI_2^{(N)}|\Psi_{i}^{(+)}\rangle
\nonumber\\&\approx&\langle\phi_f|\phi_f^{(N)}\rangle
\langle\bi{k}_f\phi_f^{(N)}|T|\phi_i^{(N)}\bi{k}_i\rangle,
\label{ccnoex}
\end{eqnarray}
where the $N$-state $T$ matrix is obtained from \eref{TLS}, and 
the states $\phi_n^{(N)}$ have been obtained in such a way that 
given a particular eigenstate $\phi_f$ of energy $\epsilon_f$
(discrete or continuous), for some $n=f$ we have
$\epsilon_f^{(N)}=\epsilon_f$, and hence
\begin{equation}
\langle\phi_f|\phi_n^{(N)}\rangle\approx\delta_{fn}
\langle\phi_f|\phi_f^{(N)}\rangle.
\end{equation}

For discrete $\epsilon_f<0$ we need $N$ to be sufficiently large so that
$\langle\phi_f|\phi_f^{(N)}\rangle\approx1$ and
$\langle\phi_i|\phi_i^{(N)}\rangle\approx1$. In this case we use the
$T$ matrix calculated in \eref{TLS} directly. For $\epsilon_f>0$ with
$\langle\phi_f|\equiv\langle\bi{q}_f^{(-)}|$ the
$T$ matrix in \eref{TLS} is multiplied by the overlap
$\langle\bi{q}_f^{(-)}|\phi_f^{(N)}\rangle$, which has the effect of
restoring the continuum boundary conditions and introduces a
one-electron Coulomb phase.

The close-coupling without exchange $N$-state approximation to the
experimentally measured TDCS is
\begin{eqnarray}
\frac{d^3\sigma^{(N)}}{d\Omega_1d\Omega_2dE_2}&=&
|\langle\bi{q}^{(-)}_f|\phi_f^{(N)}\rangle\langle \bi{
k}_f\phi^{(N)}_f|T|\phi_i^{(N)}\bi{
k}_i\rangle|^2\nonumber\\
&&+|\langle\bi{k}^{(-)}_{f'}|\phi_{f'}^{(N)}\rangle
\langle \bi{q}_{f'}\phi_{f'}^{(N)}|T|\phi_i^{(N)}\bi{k}_i\rangle|^2.
\label{CCCnoE}
\end{eqnarray}
This is a generalisation of \eref{bornc2}. With such a definition the
SDCS is symmetric about $E/2$ and the total ionization cross section
would be obtained by integration to $E/2$.
It is helpful to think of the second term in \eref{TLS} as a second
order correction to the Born approximation. As such, it vanishes at
high energies leaving just the Born approximation for both the
discrete excitation and ionizing collisions. Numerically, we find
$\langle\bi{q}^{(-)}_f|\phi_f^{(N)}\rangle
\langle \bi{k}_f\phi^{(N)}_f|V|\phi_i^{(N)}\bi{k}_i\rangle\approx
\langle \bi{k}_f\bi{q}^{(-)}_f|V|\phi_i^{(N)}\bi{k}_i\rangle$ to a
high accuracy due to the short-ranged $\phi_i^{(N)}$ negating the
long-ranged behaviour of $\bi{q}^{(-)}_f$.

For unequal energy-sharing the two terms in \eref{CCCnoE} are very
different and 
converge to their respective Born approximations with increasing
energy. In this case the first term converges to the true scattering
amplitude while the second converges to zero.

Note that for equal energy-sharing $f=f'$, but
the two terms are still generally different owing to the vector
nature of momenta. They are equal to each other for the so-called
coplanar doubly symmetric ($E_A=E_B$ and $\theta_A=-\theta_B$)
geometry. However, while exchange is neglected this approximation will 
not work well for this special case, and has only value whenever the
SDCS at $E/2$ is very much smaller than for the highly asymmetric
energy-sharing.

\subsection{The close-coupling with exchange approximation}
In the momentum-space formulated close-coupling equations \eref{TLS}
introduction  
of exchange results in a simple modification of the interaction
potential $V$ by
$V_S=V+(-1)^S(H-E)P_r$, where $P_r$ is the space exchange operator
\cite{BS92}. We then solve
\begin{eqnarray}
\fl
\langle\bi{k}_f\phi_f^{(N)}|T_S|\phi_i^{(N)}\bi{k}_i\rangle 
&=&
\langle\bi{k}_f\phi_f^{(N)}|V_S|\phi_i^{(N)}\bi{k}_i\rangle\nonumber\\ 
&&+\sum_{n=1}^N\int d^3k\frac{\langle\bi{
k}_f\phi_f^{(N)}|V_S|\phi_n^{(N)}\bi{k}\rangle 
\langle\bi{k}\phi_n^{(N)}|T_S|\phi_i^{(N)}\bi{k}_i\rangle}
{E+i0-\epsilon_n^{(N)}-k^2/2}
\label{TLS(S)}
\end{eqnarray}
separately for $S=0,1$. Subsequently, the $S$-dependent differential
cross sections are obtained using \eref{CCCnoE} i.e.
\begin{eqnarray}
\frac{d^3\sigma_S^{(N)}}{d\Omega_1d\Omega_2dE_2}&=&
|\langle\bi{q}^{(-)}_f|\phi_f^{(N)}\rangle\langle \bi{
k}_f\phi^{(N)}_f|T_S|\phi_i^{(N)}\bi{
k}_i\rangle|^2\nonumber\\
&&+|\langle\bi{k}^{(-)}_{f'}|\phi_{f'}^{(N)}\rangle
\langle \bi{q}_{f'}\phi_{f'}^{(N)}|T_S|\phi_i^{(N)}\bi{k}_i\rangle|^2,
\label{CCC}
\end{eqnarray}
and the CCC-calculated spin-averaged cross section for e-H ionization
is evaluated as
\begin{equation}
\frac{d^3\sigma^{(N)}}{d\Omega_1d\Omega_2dE_2} =
\frac{1}{4}\frac{d^3\sigma_0^{(N)}}{d\Omega_1d\Omega_2dE_2}
+\frac{3}{4}\frac{d^3\sigma_1^{(N)}}{d\Omega_1d\Omega_2dE_2}.
\end{equation}

The close-coupling with exchange approximation is equivalent to
\begin{equation}
|\Psi_{Si}^{(+)}\rangle\approx (1+(-1)^S
P_r)\sum_{n=1}^N|\phi_n^{(N)}f^{(N+)}_{Sni}\rangle. 
\end{equation}
Thus, the total wave function is antisymmetric in all space of the two 
electrons, but is zero when both $r_1$ and $r_2$ are large. 

The change from $V$ to $V_S$ is not entirely trivial. There are
extra computational difficulties due to non-uniqueness problems,
but these have been dealt with adequately, see
\citeasnoun{BS92} for details.

With increasing total energy $E$ the contribution of the exchange part 
of $V_S$ diminishes faster than the direct part $V$. Eventually we may 
totally drop exchange to obtain \eref{TLS}, and with further increase
of energy obtain the Born approximation. 

Introducing exchange to the close-coupling formalism does not result
in the scattering amplitudes obeying the symmetrization postulate
\eref{sympos}. In particular, 
\begin{equation}
\langle\bi{q}^{(-)}_f|\phi_f^{(N)}\rangle\langle \bi{
k}_f\phi^{(N)}_f|T_S|\phi_i^{(N)}\bi{
k}_i\rangle\ne(-1)^S
\langle\bi{k}^{(-)}_{f'}|\phi_{f'}^{(N)}\rangle
\langle \bi{q}_{f'}\phi_{f'}^{(N)}|T_S|\phi_i^{(N)}\bi{k}_i\rangle
\label{symcon}
\end{equation}
generally. Note that though $\bi{q}_{f'}=\bi{q}_{f}$ and $\bi{
k}_{f'}=\bi{k}_f$ the two states $\phi^{(N)}_f$ and $\phi^{(N)}_{f'}$ 
are very different for $f\ne f'$. \citeasnoun{BC99} claim (Eq.(20) of
their paper says that the CCC amplitudes converge to the true
amplitudes) that in the limit of
infinite $N$ there should be equality in \eref{symcon}, and hence
double counting of the ionization cross sections. While we are
unable to perform such calculations, the fact that the two terms
converge to their respective Born estimates with increasing energy
indicates that their derivation is in error. The source of the error
we suspect to be in the way the limit $N\to\infty$ is taken ignoring
how this affects the close-coupling boundary conditions.
What we do observe, with
increasing $N$, is that the close-coupling with exchange calculations yield 
diminishing cross sections for amplitudes $\langle \bi{
q}_{f'}\phi_{f'}^{(N)}|T_S|\phi_i^{(N)}\bi{k}_i\rangle$, where
$q_{f'}^2/2 < \epsilon_{f'}^{(N)}$. This has led to the suggestion
that the CCC calculations should converge to a step-function SDCS
\cite{B97l}, with numerical problems arising whenever the size of the
step at $E/2$ is substantial. This idea has gained further support
from \citeasnoun{MKW99} and \citeasnoun{BRIM99} who studied the e-H S-wave
model. Unfortunately, a mathematical proof is still lacking.

Introduction of exchange removes the distinguishability between the
two electrons of energy $\epsilon_{f}^{(N)}$ and
$E-\epsilon_{f}^{(N)}$ for a particular ionization process, however
this process is still calculated twice: once with the electron of
energy $\epsilon_{f}^{(N)}$ being treated by a pseudostate and once as 
a plane wave. It is these two treatments of a single ionization process 
that are theoretically distinguishable. The step-function idea says
that for infinite $N$ one of these is zero.
\subsection{Equal energy-sharing kinematics}
Most recently \citeasnoun{S99l} has made substantial progress in the
understanding of the problem. By also studying the S-wave model he
deduced that at equal-energy-sharing the true ionization amplitudes
$f_S(\bi{k},\bi{q})$  may be deduced from those
obtained in the close-coupling theory $f_S^{(N)}(\bi{k},\bi{q})$ by
\begin{equation} 
f_S(\bi{k},\bi{q})=f_S^{(N)}(\bi{k},\bi{q})+(-1)^S f_S^{(N)}(\bi{
q},\bi{k}).
\label{ampA}
\end{equation}
Consequently, he concluded that the CCC-calculated $k=q$
amplitudes did converge with increasing $N$, but to half the true
scattering amplitude, at least in the considered model. The cross
sections are obtained from $|f_S(\bi{k},\bi{q})|^2$ as opposed to
the integral preserving estimate we suggested $|f_S^{(N)}(\bi{k},\bi{
q})|^2 + |f_S^{(N)}(\bi{q},\bi{k})|^2$. Can the two prescriptions be 
reconciled? 

Firstly, given the observation of
\citeasnoun{S99l} that the CCC amplitudes converge to half the true
amplitudes for the model problem we suppose this is also the case for
the full problem and so our prescription yields cross sections a
factor of two too low in all equal-energy-sharing cases. This has been
previously observed in the case of e-He ionization at 64.6~eV
\cite{BFRE97_64}, 44.6~eV \cite{Rioual98} and 32.6~eV
\cite{BFRE98}. Thus, only for equal-energy-sharing we should have an
extra factor of two multiplying the incoherent combination of the
$f_S^{(N)}$. Being only at a single point this doesn't affect the
integral that leads to the correct total ionization cross section.

Now, we have noted earlier \cite{BFRE97_64} that the two
terms $|f_S^{(N)}(\bi{k},\bi{
q})|^2$ and $|f_S^{(N)}(\bi{q},\bi{k})|^2$ are substantially
different and are necessary together to yield accurate angular
distributions. If we write
\begin{equation}
f_S^{(N)}(\bi{k},\bi{q})=(-1)^Sf_S^{(N)}(\bi{q},\bi{
k})+\delta_S^{(N)}(\bi{k},\bi{q}),
\label{CCCampsym}
\end{equation}
where $\delta_S^{(N)}$ is some (small) number. Then
\begin{eqnarray}
\label{approx}
2(|f_S^{(N)}(\bi{k},\bi{
q})|^2 + |f_S^{(N)}(\bi{q},\bi{k})|^2)&=&
|f_S^{(N)}(\bi{k},\bi{
q}) + (-1)^Sf_S^{(N)}(\bi{q},\bi{k})|^2\\
&&-|\delta_S^{(N)}(\bi{k},\bi{q})|^2\nonumber,
\end{eqnarray}
whereas the difference between $|f_S^{(N)}(\bi{
k},\bi{q})|^2$ and
$|f_S^{(N)}(\bi{k},\bi{q})-\delta_S^{(N)}(\bi{k},\bi{q})|^2$ is 
much more substantial.

Thus, the coherent and incoherent combinations of amplitudes in
\eref{approx} are effectively simply doublings. The claim
\eref{CCCampsym} (assuming small $\delta_S^{(N)}$) is a very strong
one and is far from 
obvious. Consider the CCC-calculated amplitude in partial wave form
\begin{equation}
f_{JS}^{(N)}(Lk,lq)=e^{i\sigma_l(q)}|\langle
q^{(-)}_l|\phi^{(N)}_{fl}\rangle| 
\langle k_L\phi^{(N)}_{fl}|T_{JS}|\phi^{(N)}_{i0}k_0\rangle,
\end{equation}
where $J$ is total orbital angular momentum,
$\epsilon^{(N)}_{fl}=q^2/2$ and $\sigma_l(q)$ is the full complex phase
arising from the overlap $\langle
q^{(-)}_l|\phi^{(N)}_{fl}\rangle$. Given that $k=q$ interchange of $l$
and $L$ has the effect of 
explicitly changing the phase as well as the $T$-matrix
obtained from \eref{TLS(S)}. Yet together, the resulting amplitudes
satisfy \eref{CCCampsym}. Furthermore, since $l\le l_{\rm max}$ with
$|J-l|\le L\le J+l$,
we need sufficiently large $l_{\rm max}$ that interchange of $L$ and
$l$ was possible for all substantial $T_{JS}$.
To demonstrate \eref{approx} graphically, in all of
the following figures that present equal-energy-sharing kinematics we
give both sides of \eref{approx} for the two spins.

One may ask which of the two sides of \eref{approx} is more
accurate. Unfortunately, even equality does not guarantee accuracy of
the amplitudes, only correct symmetry. In other words, satisfaction of
\eref{approx} is necessary but not sufficient. The right side of
\eref{approx} has the advantage of looking compatible with
indistinguishable treatment of the two electrons, and so being able to readily
define the 
final amplitude to be used in generating the cross sections, which
will always have the correct symmetry irrespective of what the
underlying CCC amplitude is. This is a strength
and a weakness, as 
it looses sensitivity to the accuracy of the CCC calculation. The left 
side of \eref{approx} is more sensitive, since for example, for the
doubly symmetric geometry both terms must yield zero for the triplet
case. The right side has the advantage of not requiring the
step-function idea or the combination of amplitudes at $E/2$ as a limiting 
procedure of amplitudes on either side of $E/2$. The most sensitive
test would be to simply use $2f_S^{(N)}(\bi{k},\bi{q})$ or
$2f_S^{(N)}(\bi{q},\bi{k})$ as the amplitudes. The factor of
two is due to the convergence to half the true magnitude at $E/2$.

We should mention that the demonstration of failure of a coherent
combination of amplitudes for e-He equal-energy-sharing ionization at 64.6~eV
\cite{BFRE97_64} was due  
to the fact that in the case of helium the correct coherent
combination is more complicated, and requires derivation along the
lines given by \citeasnoun{S99l} for the e-H system.
The one given, $\sum_{s=0,1}|F^{(N)}_s(\bi{k},\bi{q})+F_s^{(N)}(\bi{q},\bi{
k})|^2$, where $s$ is the spin of the frozen-core two-electron continuum
wave, yields the wrong answer. The intuitive combination 
$|F^{(N)}_0(\bi{k},\bi{q})+F_1^{(N)}(\bi{q},\bi{
k})|^2+ |F^{(N)}_1(\bi{k},\bi{q})+F_0^{(N)}(\bi{q},\bi{
k})|^2$ yields a factor of two difference from the incoherent
combination used, but requires formal derivation.

\subsection{Asymmetric energy-sharing kinematics}
What about asymmetric energy-sharing? \citeasnoun{S99l} shows that
different logarithmic phases on either side of $E/2$ lead to
difficulty in defining the ionization amplitudes, unless the
CCC-calculated amplitude for $q>k$ was identically zero.
Though we are as yet unable to prove analytically the step-function
hypothesis, which \citeasnoun{BC99} believe to have proved to be
incorrect, all of our numerical evidence is consistent with it. Certainly, 
for the purpose of making comparison with experiment, it holds in our
finite calculations for the substantially asymmetric excess
energy-sharing kinematics. In these cases the second term in \eref{CCC}
is insignificant compared to the first. Does the first term yield the
true scattering amplitude? When convergent, as it is at high energies
\cite{BF96}, we suspect so. This is also implied by the analysis of
\citeasnoun{S99l}. 
At sufficiently low energies we find that
convergence to a desirable accuracy is unable to be obtained for the
SDCS with $q<k$. Our choices are then to present results as they are,
or attempt to estimate what the true SDCS should be and rescale the
CCC-calculated TDCS to this SDCS. 

How can we estimate what the true SDCS might be? Fortunately, the
underlying physics suggests that the functional form of the SDCS is
likely to be simple and, at sufficiently low energies, may be modelled
relatively accurately by a quadratic. We already know the integral
accurately ab initio, and the point of symmetry, requiring just one
more parameter to fix the quadratic. To do so \citeasnoun{BFRE98} have
observed that the value of the e-He SDCS at zero secondary energy was quite 
stable and used this to fix the estimate of the true SDCS. The
resultant rescaling lead to a factor of approximately two increase in
the CCC-calculated TDCS at equal-energy-sharing, and good
agreement with absolute experiment. A similar idea was
used for the e-H system at 15.6~eV \cite{B99jpbl}, resulting in a factor of 2.7
increase, still a further factor of two below experimental absolute
value determination. 

The oscillations in the CCC-calculated SDCS have been well-documented, and
we have been unable to explain, until now, apparent
convergence of the CCC results at equal-energy-sharing
\cite{BFRE97_64,REBF97,B97l,Rioual98,B99jpbl}. 
Now, thanks to the analysis of
\citeasnoun{S99l}, we know the value of the true SDCS at $E/2$.
This SDCS, calculated according to
\eref{intE} or \eref{intE/2}, is a factor of four or two lower than
the true SDCS, respectively. This has been tested by
comparison with the benchmark SDCS calculations
of \citeasnoun{BRIM99} as reported by \citeasnoun{S99l}. 

Another strong test of this idea is found by consideration of double
photoionization (DPI). Here the CCC method has yielded accurate total
\cite{KB98_2}  and differential \cite{KB98jpbl,Braeuning98} ionization 
cross sections. The rescaling of the CCC TDCS relied on the work of
\citeasnoun{PS95} who gave demonstrably accurate estimates of the
total cross sections, and arguably equally accurate estimates of the
SDCS($E/2$), from 2 to 80~eV above threshold. This is
particularly helpful for us as it allows a thorough comparison of the
CCC-calculated SDCS($E/2$). We have performed this check for the
published CCC-calculated DPI SDCS, and at the excess energies
presented here,  and find a
factor of two difference, generally to within 5\%.

Accordingly, rather than assuming a stable SDCS(0) derived from
examination of near threshold total ionization cross sections, we fix
the third parameter of the quadratic estimate of the SDCS by the
estimate of the true SDCS($E/2$) obtained from the raw CCC-calculated
SDCS($E/2$) multiplied by four, i.e the same value as obtained from
either side of \eref{approx}.
This is particularly helpful in the present e-H ionization case, where 
the close-coupling equations are solved separately for the two total
spins. For each total spin rather than attempting to estimate SDCS(0)
we simply obtain SDCS($E/2$) directly from the CCC calculations, 
and hence the quadratic SDCS estimate.

\section{Results and Discussion}
\begin{figure}
\epsffile{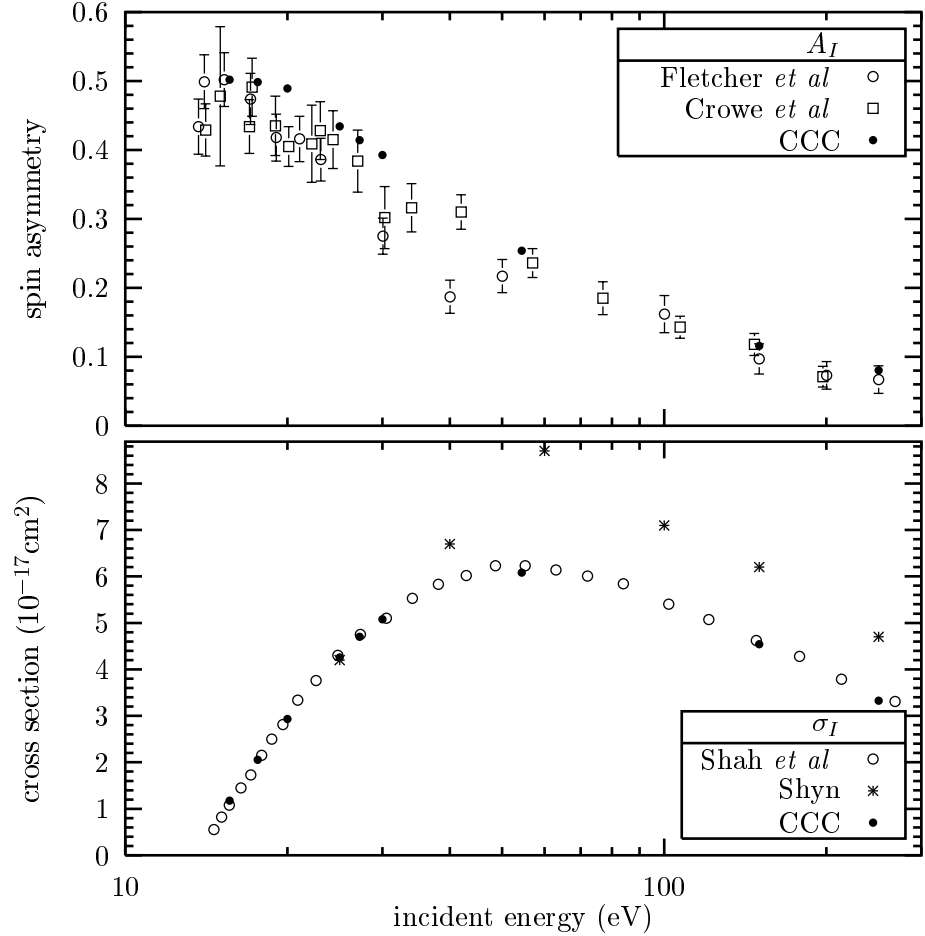}
\caption{Total ionization cross section $\sigma_I$ and its spin
asymmetry $A_I$ as a function of energy. The present results are
denoted by CCC, the measurements are due to
\protect\citeasnoun{SEG87},
\protect\citeasnoun{Cetal}, \protect\citeasnoun{Fetal} and
\protect\citeasnoun{Shyn92}. 
\label{ics}
} \end{figure}

Before looking at the detailed results of the individual energies
considered, we present in \fref{ics}  the total ionization cross section
(TICS) and its spin asymmetry as
a function of energy. The CCC calculations 
at the individual energies (solid dots) will be detailed later.
We see excellent agreement between the CCC calculations and the
experiment, with the exception of the data of \citeasnoun{Shyn92}. The
experimental technique of \citeasnoun{SEG87} is specifically aimed at the
total ionization cross section, whereas \citeasnoun{Shyn92} obtained
it after a double integration of doubly differential cross
section (DDCS) measurements. Good agreement with the spin asymmetries
indicates correct spin-dependent total ionization cross sections at
all energies.
The quality of agreement between theory and experiment was first
presented by \citeasnoun{BS93l}. Since that time other close-coupling
methods have also obtained similar results \cite{KW95,BB96H,SBBB97}.

The utility of the CCC calculations depends on obtaining
convergence with increasing $N=\sum_lN_l$. This means convergence with
target-space angular 
momentum $l_{\rm max}$ and number of states $N_l$ within each $l$. We
take $N_l=N_0-l$ as this leads to a similar integration rule in the
continuum for each $l$, of importance at low energies
\cite{B99jpbl}. This allows convenient
labelling of the calculations by CCC($N_0,l_{\rm max}$). All of the
calculations performed required substantial computational
resources. The higher energy calculations required around 1G of RAM,
while the lower energy ones required up to 2G of RAM.

At high enough 
energies most theories, those that satisfy the symmetrization
postulate, and those that don't, yield much the same results for highly 
asymmetric energy-sharing kinematics. We wish to demonstrate
that the CCC differential cross sections as defined in \eref{CCC} and
\eref{CCCnoE} also do so.

\subsection{Incident electron energy 250 eV}
We begin our study with $E_0$=250~eV. In performing the calculations we 
need to be mindful of which experiment we wish to describe. The
experiment of \citeasnoun{EJKS86} has $E_B$=5~eV, and so we ensure, by 
varying the Laguerre exponential fall-off parameter $\lambda_l$
\cite{BS92}, that one of the states $\phi_{nl}^{(N)}$ had the energy
$\epsilon_{nl}^{(N)}$=5~eV. A number of CCC($N_0,l_{\rm max}$)
calculations were performed, but we present the results from only the
biggest, CCC(15,5), which couples a total of 75 states.

\begin{figure}
\hskip2truecm\epsffile{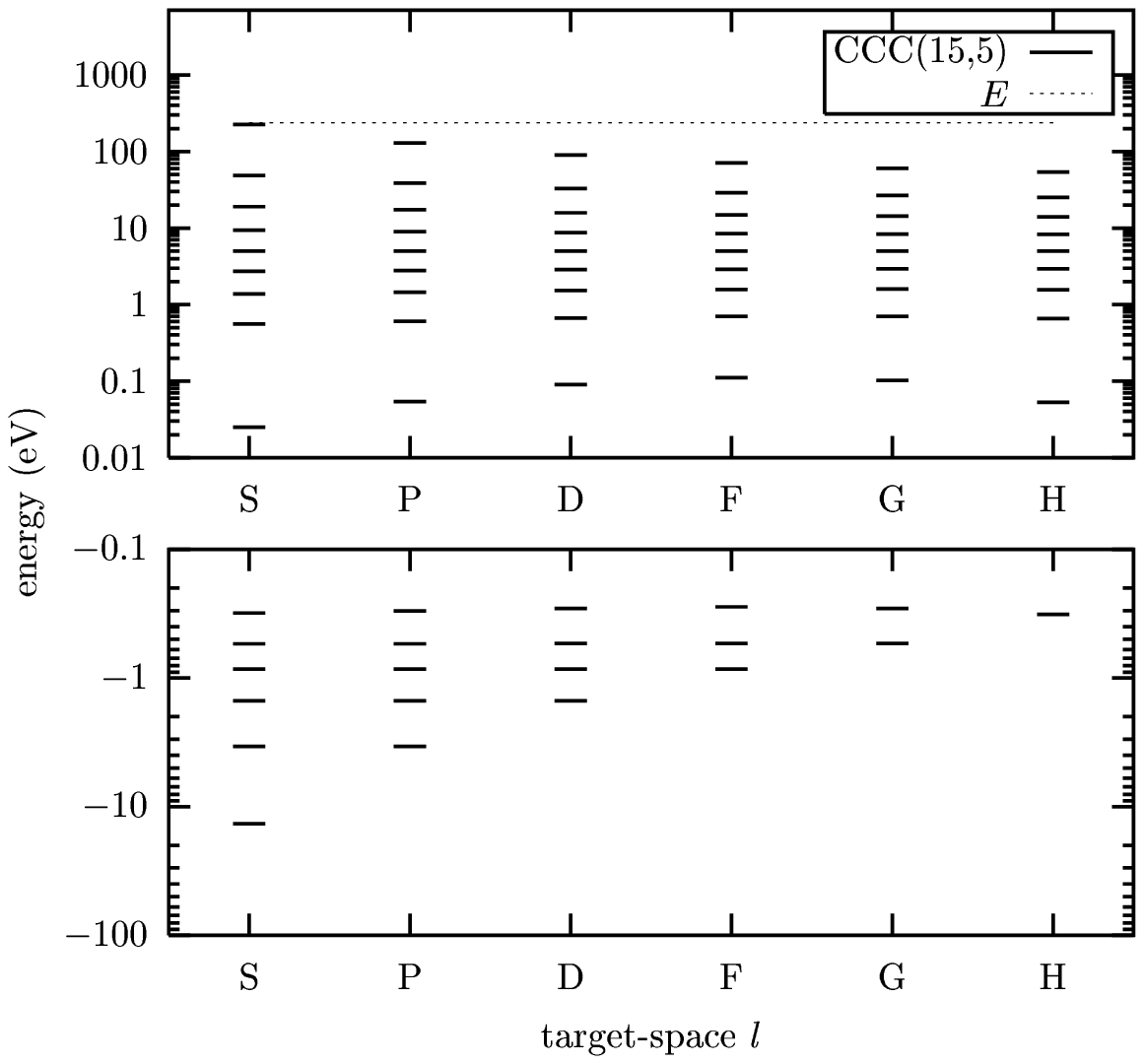}
\caption{The energy levels $\epsilon^{(N)}_{nl}$ arising in the 250~eV
e-H calculation using the CCC(15,5) model with
$\lambda_l\approx1.0$. The $\lambda_l$ were chosen so that for each
$l$ one energy was 5~eV.
\label{250en}
}\end{figure}

The energy levels of the CCC(15,5) calculation are given in \fref{250en}.
We see that the choice of states has lead to a systematic treatment of 
both the discrete and the continuous spectrum. Negative-energy states with
$n\le6$ have arisen. The $n\le5$ are good eigenstates, with the $n=6$ 
states taking into account all true $n\ge6$ discrete eigenstates. The
positive energies are approximately similarly spaced for each $l$,
particularly in the region of 5~eV. 
The total energy $E=250-13.6$~eV is greater than all
of the state energies, and hence all channels are open. The energy
levels increase approximately exponentially, and so the energy region
$[0,E/2]$ is much more densely covered than $[E/2,E]$.

\begin{figure}
\hskip0truecm\epsffile{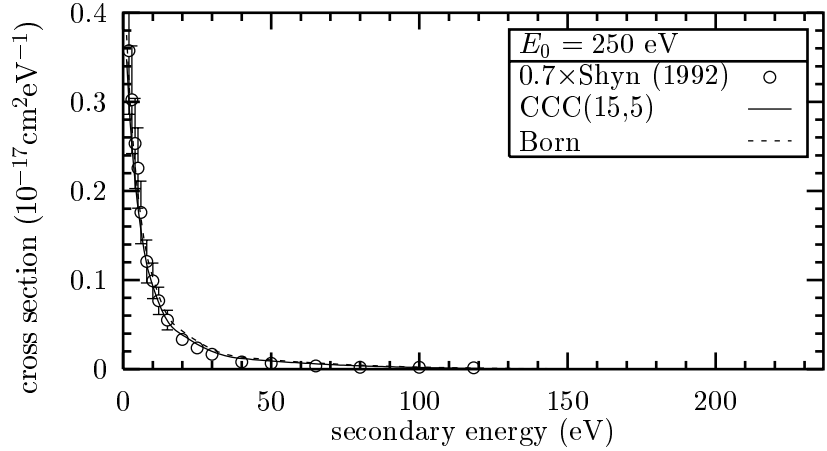}
\caption{The singly differential cross section for 250~eV
electron-impact ionization of the ground state of atomic hydrogen. The 
data of \protect \citeasnoun{Shyn92} have been scaled for consistency with the
data of \protect\citeasnoun{SEG87}, see \protect\fref{ics}.
\label{250sdcs}
}\end{figure}

In \fref{250sdcs}, we consider the SDCS arising from the CCC(15,5)
calculation. This we obtain directly from the integrated cross
sections for the excitation of the positive-energy pseudostates
\cite{BF95sdcs}, equivalent to \eref{intE}. Comparison with the data
of \citeasnoun{Shyn92} is 
given after the latter have been reduced by a factor of 0.7. This
reduction brings the experimental SDCS into consistency with the data of
\citeasnoun{SEG87}. 
There is almost no difference between the Born approximation and the
CCC(15,5) result. Both yield excellent agreement with the rescaled
experiment, though neither are symmetric about $E/2$ and hence do not
satisfy the symmetrization postulate \eref{sympos}. The theoretical
SDCS at $E/2$ is practically 
zero and remains so at higher secondary energies. The true,
experimentally measurable SDCS, would be symmetric about $E/2$, but
there is no new physics in this and does not invalidate the Born or
CCC results for the smaller secondary energies.

\begin{figure}
\vspace{0.5truecm}
\hspace{-2.5truecm}\epsffile{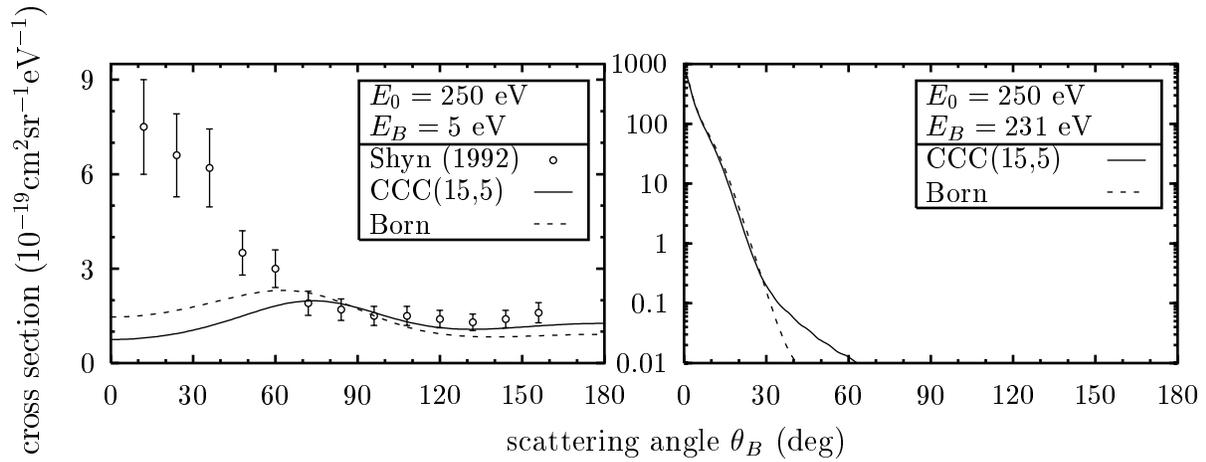}
\caption{The doubly differential cross section of the 5 and 231~eV
outgoing electrons for 250~eV
electron-impact ionization of the ground state of atomic hydrogen.
\label{250ddcs}
}
\end{figure}
The DDCS are given in \fref{250ddcs}. Unscaled data are compared with
the CCC and Born calculations. We see good agreement at the
backward angles suggesting that the experiment had some systematic
problem at the lower scattering angles. There is a little difference
between the Born and CCC calculations, but generally the two are very
similar. We also performed a CCC(15,5) calculation with no
exchange. This is indistinguishable from the presented CCC(15,5) one,
indicating that the difference with Born is due solely to
coupling. The discrepancy with experiment at forward angles is similar 
to that reported by \citeasnoun{BK93}.

\begin{figure}
\vspace{0.5truecm}
\hspace{-2.5truecm}\epsffile{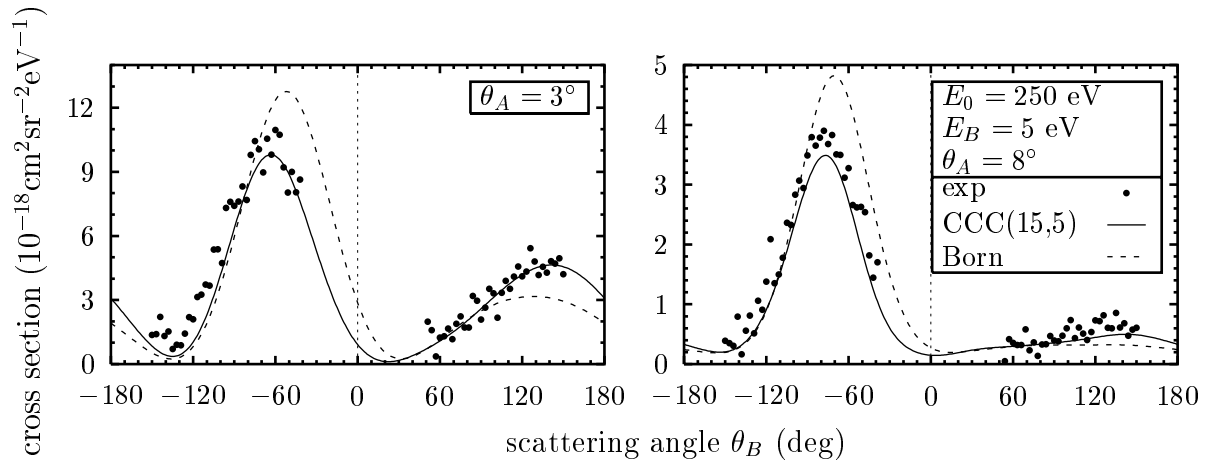}
\caption{The coplanar triply differential cross section of the
$E_B$=5~eV electron 
with the $E_A$=231~eV electron being detected at specified $\theta_A$
scattering angle for 250~eV
electron-impact ionization of the ground state of atomic hydrogen. The 
absolute measurements are due to \protect\citeasnoun{EJKS86}. Negative 
angles correspond to the opposite side of the incident beam to the
positive angles.
\label{250tdcs}
}
\end{figure}

Lastly, for this incident energy, the TDCS are presented in
\fref{250tdcs}. We see small difference between the Born and the CCC
calculation, with the latter giving complete agreement with
experiment. Comparison with the CCC(15,5) no exchange calculation,
which is pictorially indistinguishable from the presented CCC(15,5)
one, indicates that the improvement on the Born approximation is again 
solely due to coupling.

In our view, the results presented at this energy are sufficient to
invalidate the arguments of \citeasnoun{BC99}. Here the close-coupling 
formalism yields results much the same as the Born approximation and
experiment. This is not fortuitous. The second term in \eref{CCC} is
essentially zero, with the first term, in our view, having converged
to the true ionization scattering amplitudes of the problem considered.

\subsection{Incident electron energy 150 eV}
\begin{figure}
\hskip2truecm\epsffile{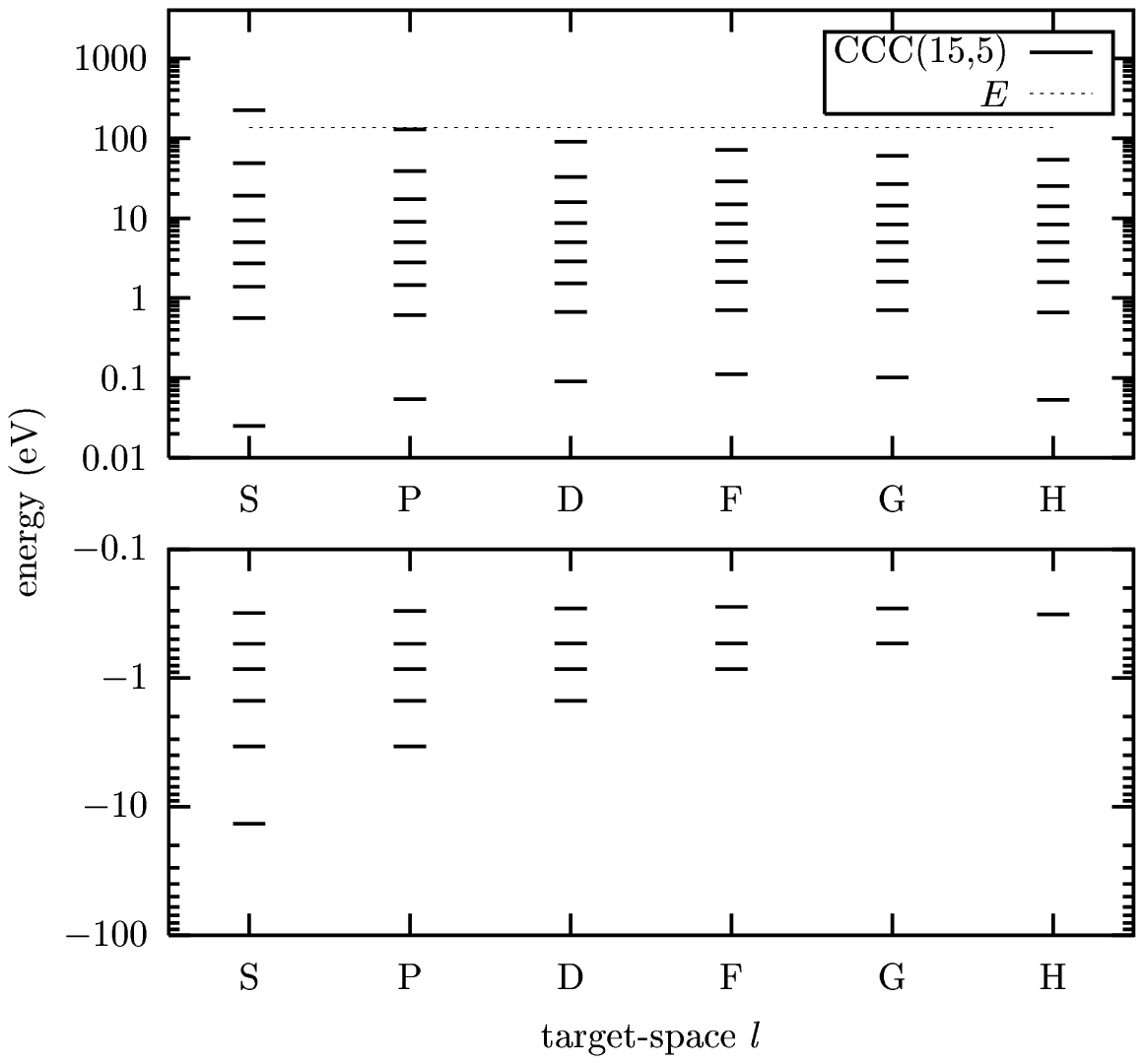}
\caption{The energy levels $\epsilon^{(N)}_{nl}$ arising in the 150~eV
e-H calculation using the CCC(15,5) model with
$\lambda_l\approx1.0$. The $\lambda_l$ were chosen so that for each
$l$ one energy was 5~eV.
\label{150en}
}\end{figure}
\begin{figure}
\epsffile{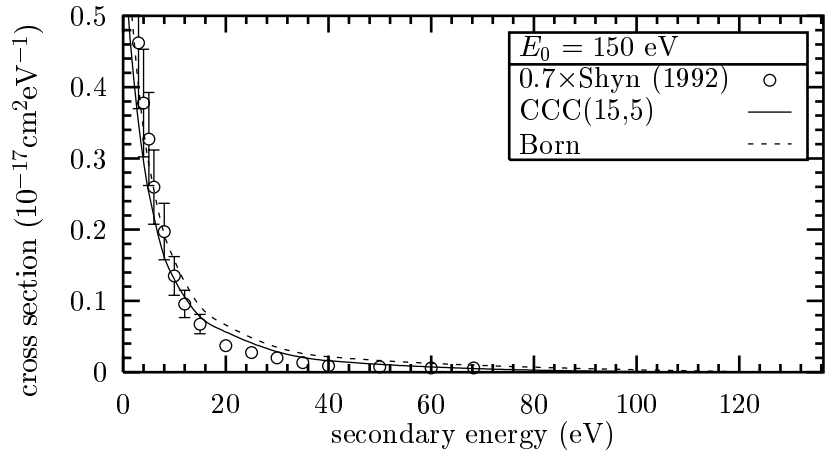}
\caption{The singly differential cross section for 150~eV
electron-impact ionization of the ground state of atomic hydrogen. The 
data of \protect\citeasnoun{Shyn92} have been scaled for consistency with the
data of \protect\citeasnoun{SEG87}, see \protect\fref{ics}.
\label{150sdcs}
}\end{figure}
We have considered e-H ionization at 150~eV in the very first
application of the CCC method to differential ionization cross
sections \cite{BKMS94}. The formalism used then varies a little from
the present in that following \citeasnoun{CW87} an attempt was previously
made to incorporate the treatment 
of higher target-space orbital angular momentum than the $l_{\rm max}$ 
used within the close-coupling equations. We no longer do so,
believing that it is more consistent to extract all of the ionization
information from only the matrix elements arising upon the solution of the
close-coupling equations.

\begin{figure}
\hspace{-3.5truecm}
\epsfbox{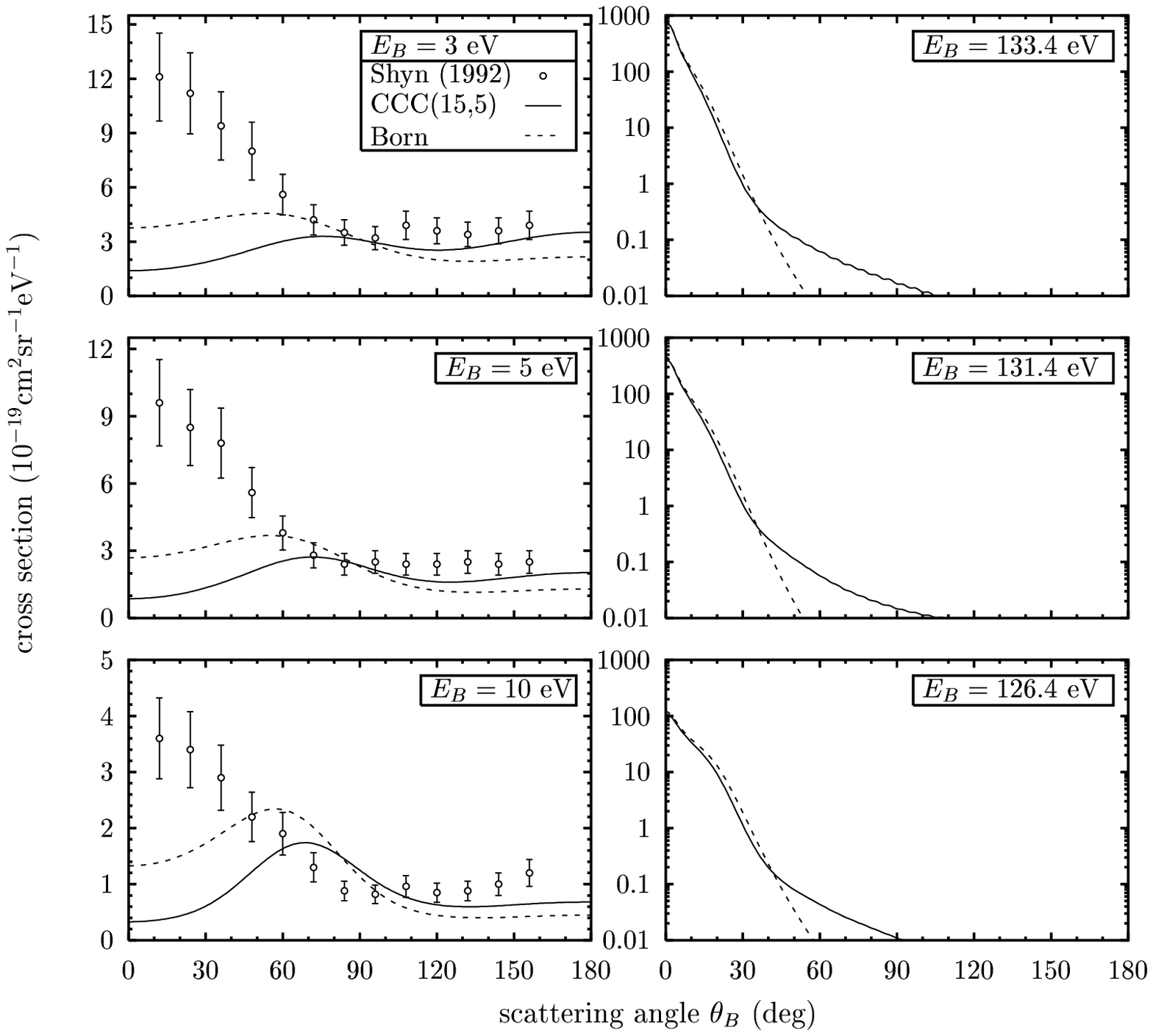}
\caption{The doubly differential cross section of the indicated 
outgoing electrons for 150~eV
electron-impact ionization of the ground state of atomic hydrogen.
\label{150ddcs}
}
\end{figure}
At this energy we have absolute experimental TDCS for three secondary
energies $E_B$=3, 5 and 10~eV \cite{EJKS86}. In a single
calculation we may vary
$\lambda_l$ to obtain only one of the $E_B$. The TDCS at other $E_B$
have to be obtained with the assistance of interpolation
\cite{BF96}. Three CCC(15,5) calculations were performed with
$\lambda_l$ varied to obtain each of the three $E_B$. Comparison of
the full set of TDCS showed little variation and so we present the
results just from the calculation where the $\lambda_l$ were  varied
to obtain $E_B=5$~eV.
The energy levels of this CCC(15,5) calculation are given in \fref{150en}.
We see that the choice of states is very similar to the case of 250~eV
incident energy
(\fref{250en}). The total energy $E=150-13.6$~eV is greater than all but
one of the state energies.

\begin{figure}
\hspace{-3truecm}\epsffile{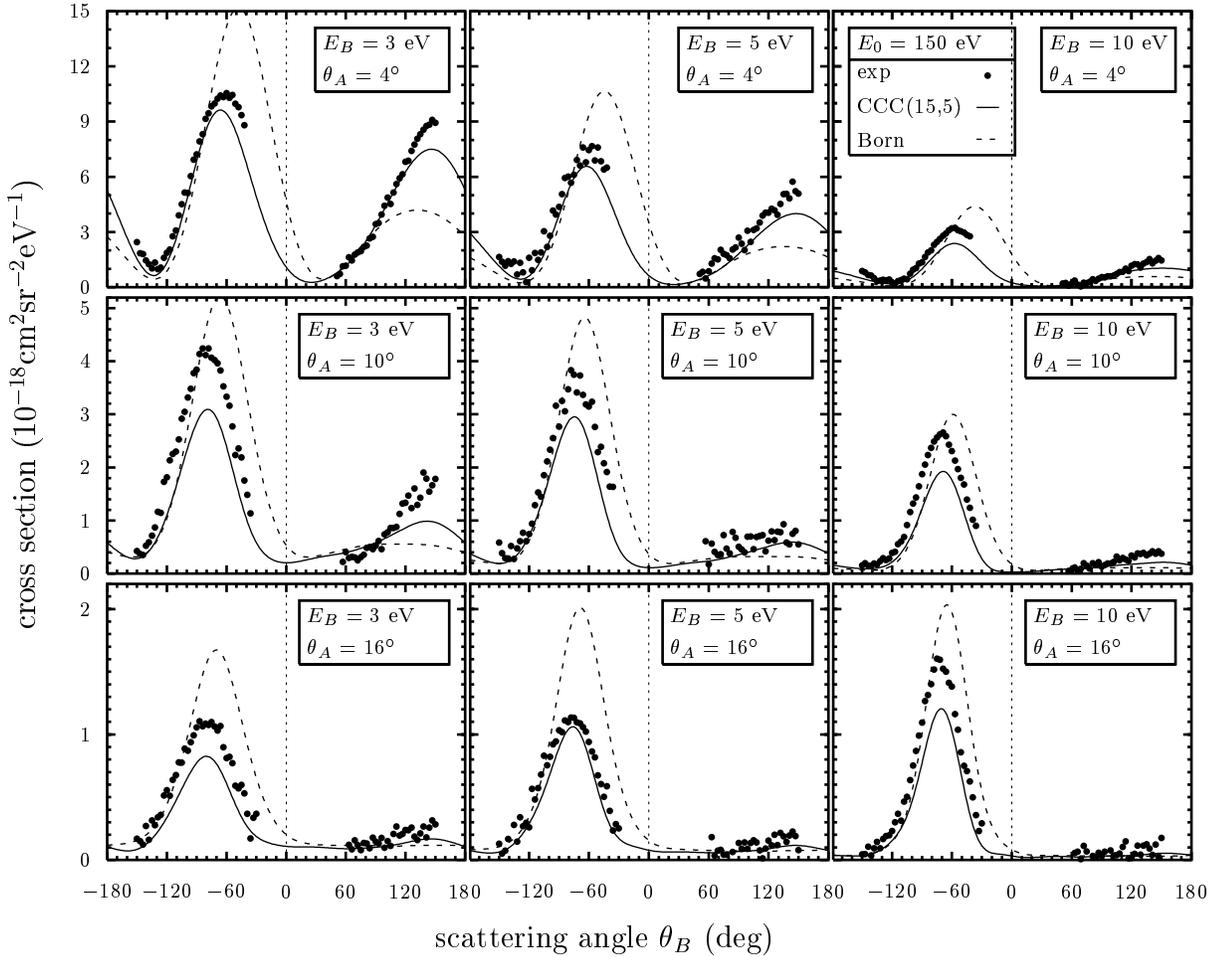}
\caption{The coplanar triply differential cross sections of the indicated
electron of energy $E_B$
with the $E_A$ electron being detected at specified $\theta_A$
scattering angle for 150~eV
electron-impact ionization of the ground state of atomic hydrogen. The 
absolute measurements are due to \protect\citeasnoun{EJKS86}.
\label{150tdcs}
}
\end{figure}
In \fref{150sdcs} the SDCS arising from the
calculation is considered. Comparison with the data of \citeasnoun{Shyn92} is
given after the latter have been reduced by again a factor of 0.7.
There is now some visible difference between the Born approximation and the
CCC(15,5) result. Again, no exchange calculations show that this is due to
neglect of coupling in the Born approximation. Both yield good
agreement with the rescaled experiment. The SDCS($E/2$) is
practically zero and hence, we suspect, there are no convergence problems.

The DDCS are given in \fref{150ddcs}. Unscaled data of
\citeasnoun{Shyn92} is compared with
the CCC and Born calculations. As one might expect the difference
between Born and CCC is somewhat bigger at this energy than at
250~eV. The smaller visible difference in the SDCS is due to the
``crossing-over''  of the two curves.
The agreement with experiment is only acceptable at intermediate and
backward angles. The fact that these
data lead to only a 30\% lower TICS than the \citeasnoun{SEG87} data is
due to the  
$\sin(\theta)$ term in the integration of the DDCS to obtain the
SDCS. 

The TDCS are presented in
\fref{150tdcs}. The difference between the Born and the CCC
calculation is quite substantial. Comparison with the CCC(15,5) no
exchange calculation indicates that the difference with the Born
approximation is primarily due to coupling.
The agreement with experiment is somewhat mixed. The fact that the
Born approximation is too high and sometimes the CCC result too low
indicates that a calculation which combines the two ideas, like a
distorted-wave Born approximation (DWBA), may occasionally yield a
better agreement with experiment than the presented CCC
calculations, see \citeasnoun{BKMS94} for some comparison with other
theory. However, we suppose that the present calculations should  
be the most accurate.

\subsection{Incident electron energy 54.4 eV}
This energy was also considered in the very first
application of the CCC method to differential ionization cross
sections \cite{BKMS94}. However, as described in the previous
subsection the formalism is now a little different, and also some
new interesting issues have since emerged.

\begin{figure}
\hskip2truecm\epsffile{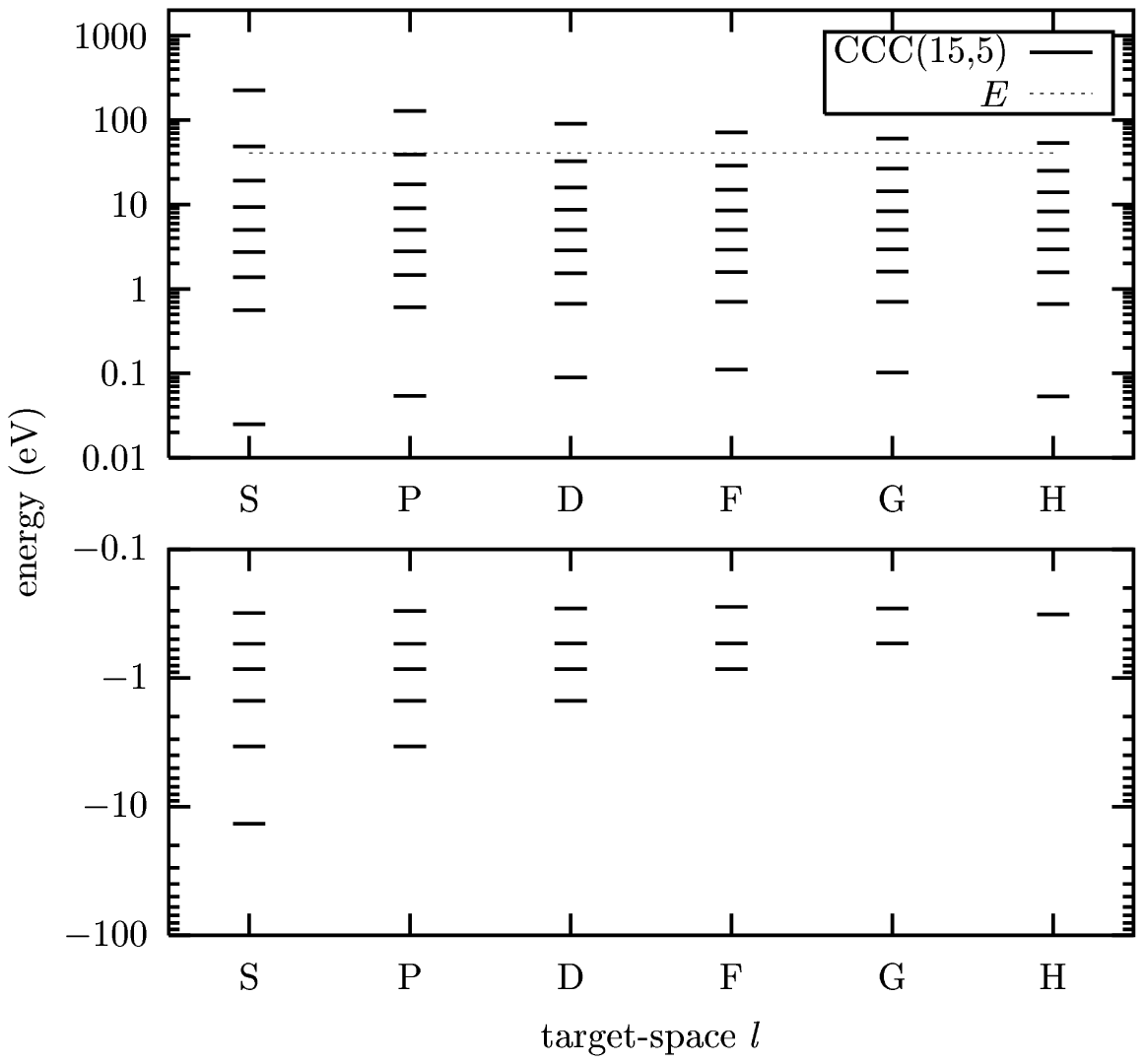}
\caption{The energy levels $\epsilon^{(N)}_{nl}$ arising in the 54.4~eV
e-H calculation using the CCC(15,5) model with
$\lambda_l\approx1.0$. The $\lambda_l$ were chosen so that for each
$l$ one energy was 5~eV.
\label{54.4en}
}\end{figure}

At 54.4~eV incident electron energy absolute experimental TDCS for 
$E_B=5$~eV exist for four angles of the fast electron \cite{Roeder96}. 
We again apply a CCC(15,5) approximation at this energy. 
The energy levels of this calculation are given in \fref{54.4en}.
The energy distribution is much the same as at 250 and 150~eV.
The total energy $E=54.4-13.6$~eV is such that there is a ``closed''
state for each $l$ (two for S-states). \citeasnoun{BC97} discussed, by
reference to the equivalent quadrature idea, the
importance of having the total energy bisect two of the
pseudothresholds. This is particularly important for small
$N_0$ and $E$. Unfortunately we are unable to have both an energy level at
5~eV and ensure that $E$ is inbetween two other energy levels. In the
present case this is not a major issue as we shall see that the SDCS
is very small at the larger secondary energies.

\begin{figure}
\epsffile{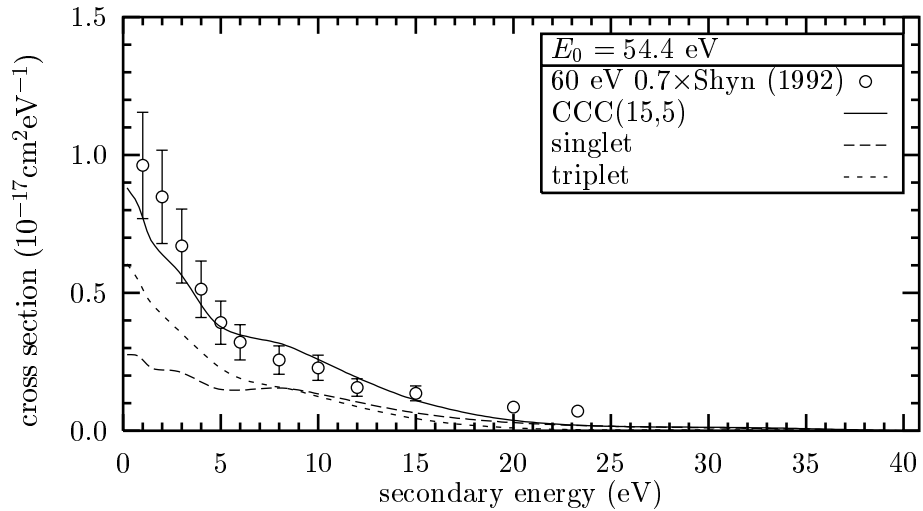}
\caption{The singly differential cross section for 54.4~eV
electron-impact ionization of the ground state of atomic hydrogen. The 
data of \protect\citeasnoun{Shyn92} have been scaled for consistency with the
data of \protect\citeasnoun{SEG87}, see \protect\fref{ics}. The
singlet and triplet contributions include the spin weights.
\label{54.4sdcs}
}\end{figure}

In \fref{54.4sdcs} the SDCS arising from the
calculation is considered with comparison of the available rescaled
60~eV data of \citeasnoun{Shyn92}.
At this energy the Born approximation is much too high and we shall
not consider it again. Instead, we shall concentrate on the importance 
of the two spin ($S=0,1$) channels. These are presented with the spin
weights included so the spin-averaged sum is simply the sum of the
singlet and triplet components. 

\begin{figure}
\hspace{-2.8truecm}\epsffile{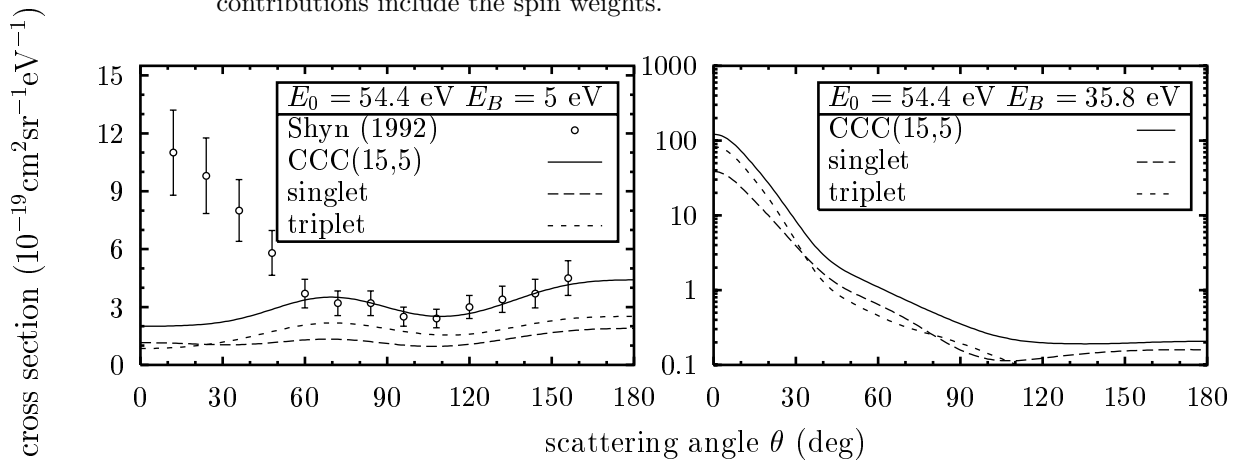}
\caption{The doubly differential cross section of the indicated 
outgoing electrons for 54.4~eV
electron-impact ionization of the ground state of atomic hydrogen. The
singlet and triplet contributions include the spin weights.
\label{54.4ddcs}
}
\end{figure}

Comparison with experiment is generally good, but looking at the
individual spin components suggests the existence of a numerical
problem. Whereas the triplet component is very smooth, the singlet one 
shows minor unphysical oscillation. It is our opinion that this is due 
to the fact the singlet SDCS at $E/2=20.4$~eV is substantially bigger
than the triplet one, which is near zero. If, as we suppose, the
step-function hypothesis \cite{B97l} is true, then the size of the
step should be relatively bigger for the singlet case. A finite
discretization of such a step function may be the cause of
the oscillation. As a consequence, there is some uncertainty in the
magnitudes of the singlet contribution at 5~eV. 
We could attempt to rescale SDCS.
However, at this energy we did not ensure an energy
point at $E/2$ for each $l$, see \fref{54.4en}. Hence
the magnitude of the SDCS($E/2$) may significantly depend on the
choice of interpolation. 

The DDCS are given in \fref{54.4ddcs}. Unscaled data of
\citeasnoun{Shyn92} are compared with
the CCC calculations. Also given are the singlet and triplet components.
The agreement with experiment is good at intermediate and
backward angles, but the systematic problem at forward angles
continues.

\begin{figure}
\hspace{-2.5truecm}\epsffile{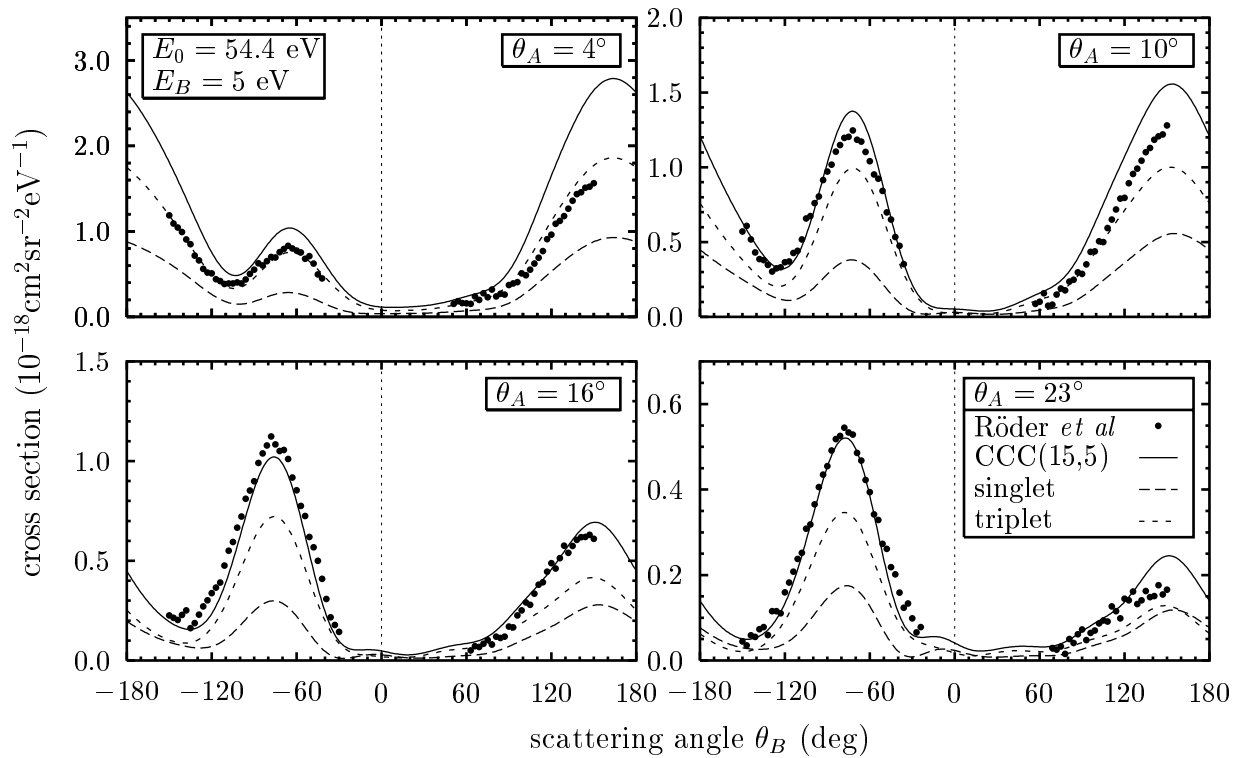}
\caption{The coplanar triply differential cross sections of $E_B=5$~eV 
electrons for 54.4~eV
electron-impact ionization of the ground state of atomic hydrogen. The 
absolute measurements are due to \protect\citeasnoun{Roeder96}.
\label{54.4tdcs}
}
\end{figure}

The TDCS are presented in
\fref{54.4tdcs}. For clarity of presentation we do not compare with
the multitude of other available theories here. Considerable comparison of 
other theories with experiment
may be found in \citeasnoun{BKMS94},  \citeasnoun{Roeder96} and
\citeasnoun{JMK97}.  
The agreement with experiment is a little disturbing for small
$\theta_A$, but improves rapidly with increasing $\theta_A$. 
Perhaps a more accurate theoretical estimate may be obtained by
marginally increasing the 
singlet component (systematically for all $\theta_A$), according to
the discussion relating to the SDCS. Looking at the data it is
difficult to argue for or against this case. What is clear is that due 
to the inherent difficulties of the CCC formalism e-H ionization and
e-He ionization have different problems in terms of comparison with
experiment. In the e-He case there is only one value of spin, here we
have two, but experiment only measures their sum.
These issues become more transparent at lower energies with equal
energy-sharing kinematics.

\subsection{Incident electron energy 30 eV}
\begin{figure}
\hskip2truecm\epsffile{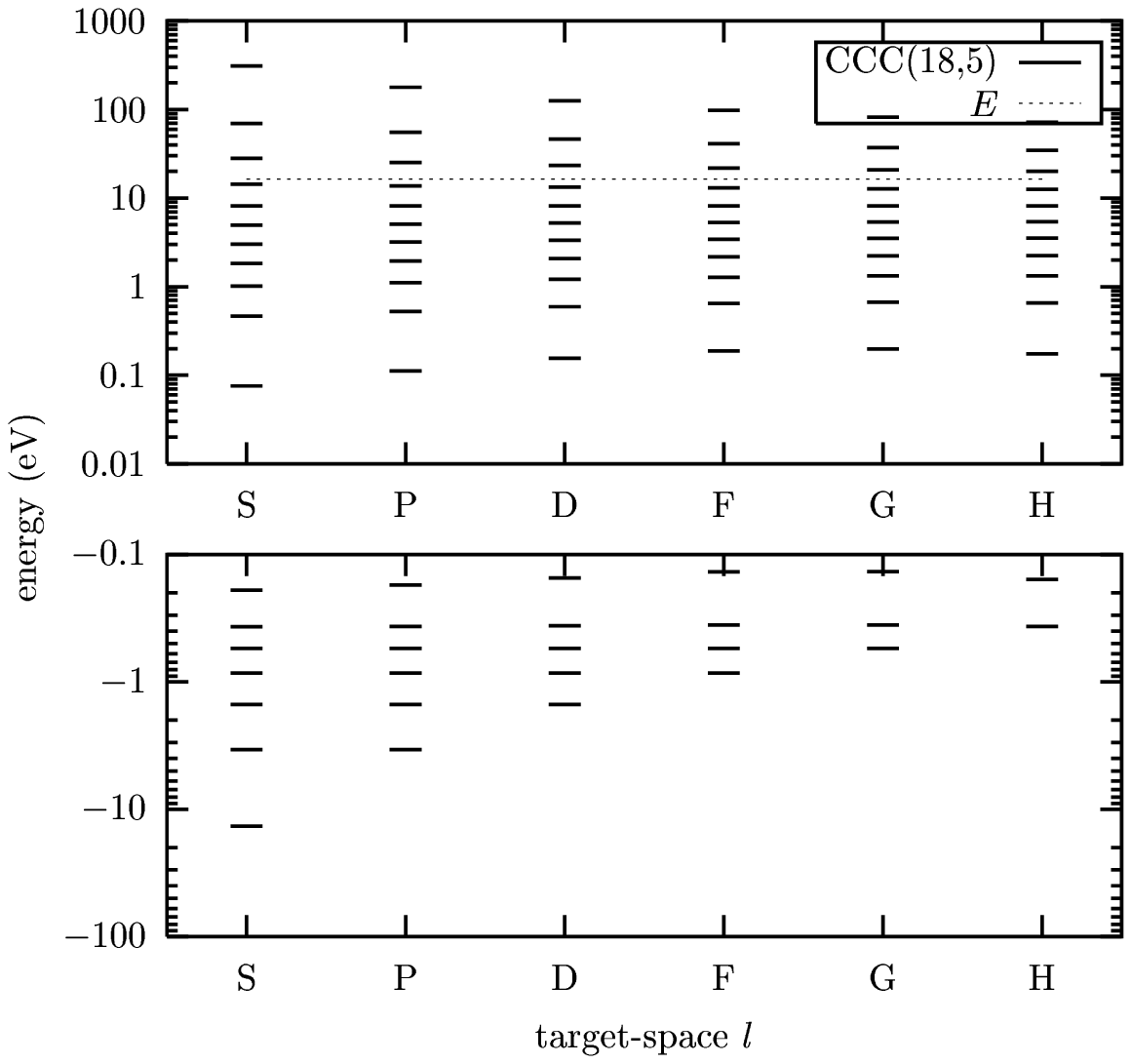}
\caption{The energy levels $\epsilon^{(N)}_{nl}$ arising in the 30~eV
e-H calculation using the CCC(18,5) model with
$\lambda_l\approx1.0$. The $\lambda_l$ were chosen so that for each
$l$ one energy was 8.2~eV.
\label{30en}
}\end{figure}

At 30~eV incident energy relative equal energy-sharing
($E_B=E_A=8.2$~eV) data exists for the coplanar fixed $\theta_{AB}$ geometries
\cite{Roeder96} and the coplanar symmetric geometry
\cite{Whelan94}. As the incident energy and hence $E$ is reduced we
need to take more care that $E$ is nearly inbetween two of the
pseudothresholds so that the integration rule associated with the open 
pseudostates ended near $E$. This issue is alleviated by having a
larger $N_0$ as then the size of the SDCS at larger secondary energies 
is further reduced. For these reasons here we present the results of a
CCC(18,5) calculation.
The energy levels of this calculation are given in \fref{30en}.
The total energy $E=16.4$~eV is such that there are three ``closed''
states for each $l$. Of the extra (over CCC(15,5)) three states for
each $l$ one has 
gone into the discrete spectrum and two into the continuum.

\begin{figure}
\epsffile{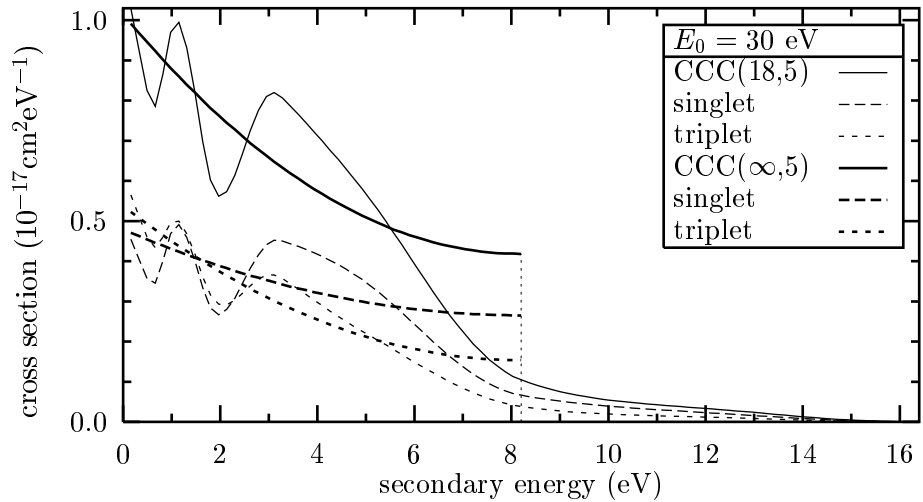}
\caption{The singly differential cross section for 30~eV
electron-impact ionization of the ground state of atomic hydrogen. 
The singlet and triplet results are obtained directly from the
CCC(18,5) calculation c.f. \eref{intE}. The
CCC($\infty,5$), singlet and triplet, curves are integral preserving
estimates with CCC$(\infty,5)=4\times$CCC(18,5) at $E/2$, see text. The 
singlet and triplet contributions include the spin weights.
\label{30sdcs}
}\end{figure}

In \fref{30sdcs} the SDCS arising from the CCC(18,5)
calculation is considered. No experimental SDCS are available at this
energy. 
At this energy the SDCS at $E/2$ is quite substantial, and thus we see 
unphysical oscillations in both the singlet and triplet
components, though integrals of both yield excellent agreement with
experiment, see \fref{ics}. The unphysical oscillations indicate
that the angular distributions will have incorrect magnitudes. We
suppose that the integral preserving quadratic estimate labelled by
CCC($\infty,5$) is the step-function that the close-coupling formalism 
would converge to for infinite $N_0$. Convergence at exactly $E/2$ is
to a quarter the height of the step, and is readily obtained in finite
calculations, as we shall see at the next considered energy. Incidentally,
the convergence in 
the SDCS with increasing $l_{\rm max}$ is particularly fast, and a
CCC(18,3) calculation gives an almost indistinguishable SDCS result.

\begin{figure}
\epsffile{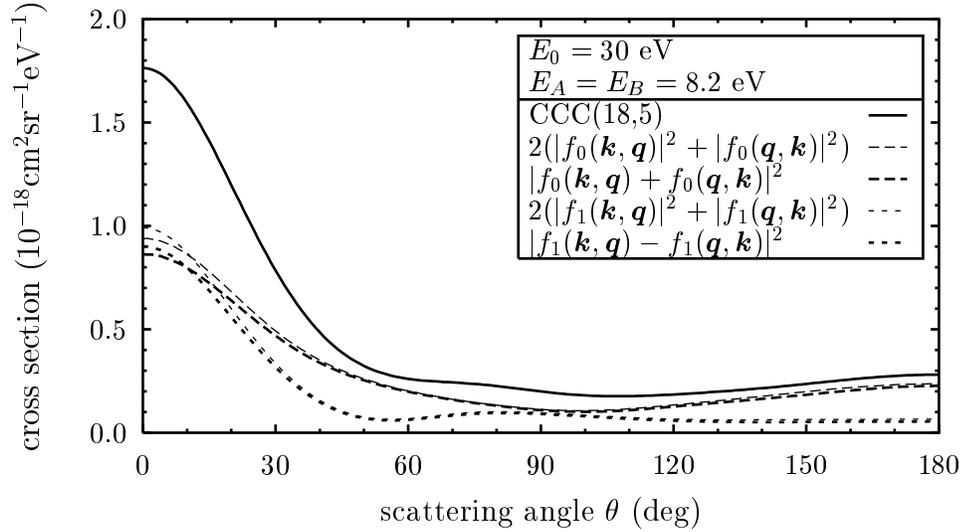}

\caption{The doubly differential cross section of the 8.2~eV
outgoing electrons for 30~eV
electron-impact ionization of the ground state of atomic hydrogen. The
singlet and triplet contributions include the spin weights, and have
been evaluated using both sides of \protect\eref{approx} prior to
integration over one of the $d\Omega$.
\label{30ddcs}
}
\end{figure}

The 30 eV DDCS, spin-weighted and the individual
singlet and triplet components, are given in \fref{30ddcs}. These are
given only for completeness as no experiment is yet available for this
case. The singlet and triplet components evaluated using both sides of 
\eref{approx} are given to show the minimal difference between the two 
prescriptions.

\begin{figure}
\hspace{-2truecm}\epsffile{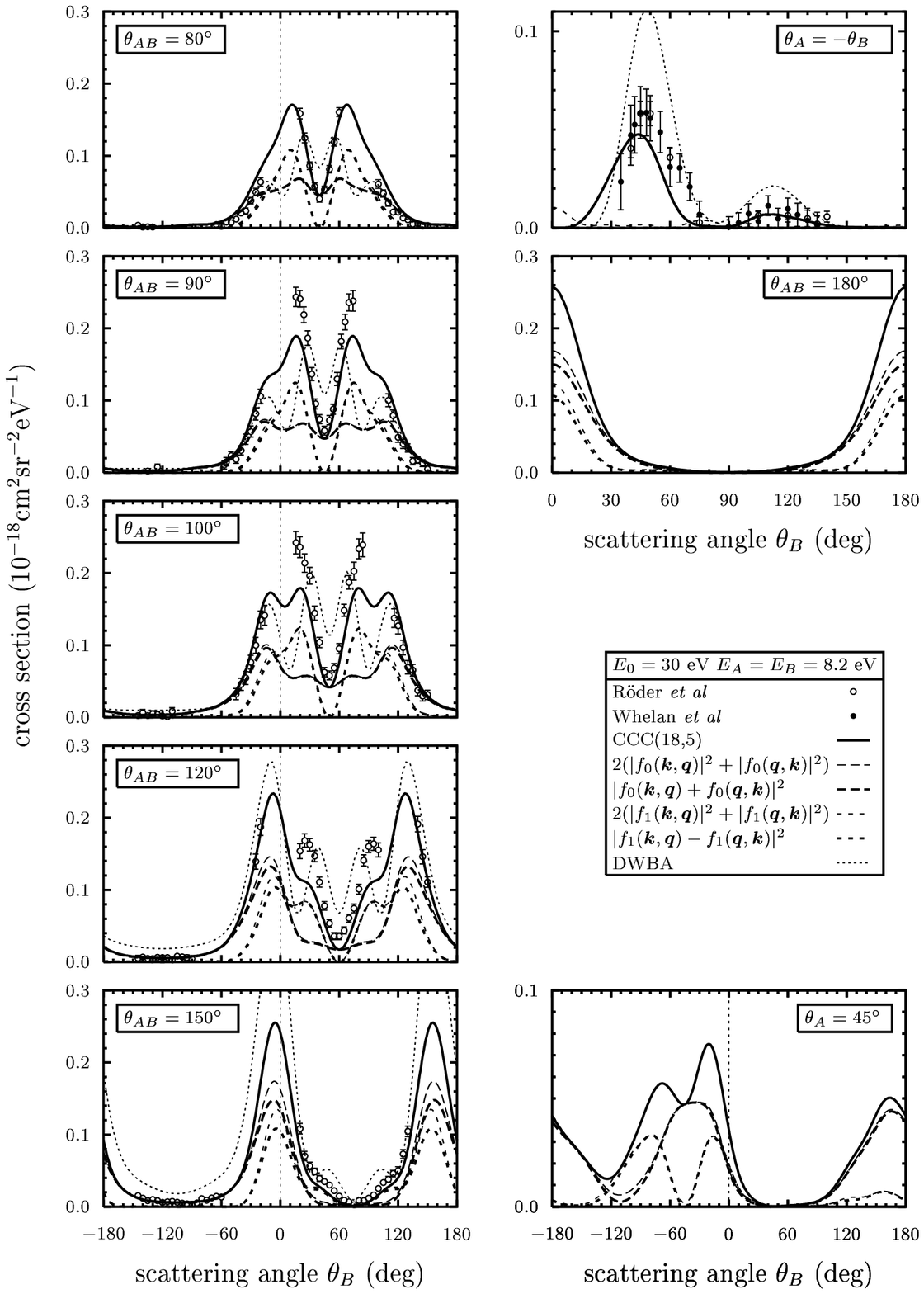}
\caption{The coplanar equal energy-sharing triply differential cross
sections  for 30~eV
electron-impact ionization of the ground state of atomic hydrogen. The 
internormalized relative $\theta_{AB}$ measurements, due to
\protect\citeasnoun{Roeder96}, have been normalised by a single factor 
to the CCC(18,5) calculation, whose singlet and triplet components are 
given according to \protect\eref{approx}. The measurements and
calculations of 
\protect\citeasnoun{Whelan94} are internormalized with those of
\protect\citeasnoun{Roeder96}.
\label{30tdcs}
}
\end{figure}

The corresponding TDCS are presented in
\fref{30tdcs}. The coplanar relative $\theta_{AB}$ measurements of
\citeasnoun{Roeder96} have been 
scaled by a single 
factor for best overall visual fit. The DWBA with polarization and
PCI effects calculation, presented in arbitrary units by
\citeasnoun{Roeder96}, has been scaled to fit experiment as done by
\citeasnoun{Roeder96}. In order to internormalize the coplanar
symmetric data presented by \citeasnoun{Whelan94} we have extracted
the symmetric geometry points from the $\theta_{AB}$ measurements. The 
symmetric geometry calculation of \citeasnoun{Whelan94} is
internormalized to the 
$\theta_{AB}$ calculations, and is the reason why it is substantially higher
than experiment compared to the initial presentation \cite{Whelan94}.

The first thing to note is the excellent agreement between the
coherent and incoherent combinations of amplitudes for both spins. The 
corresponding thick and thin curves are almost
indistinguishable. There are some examples where the difference is
quite visible. For the $\theta_{AB}=150^\circ$ case around $0^\circ$
and $160^\circ$ there is approximately 15\% difference. However, the
difference between the $|f_S(\bi{k},\bi{q})|^2$ and $|f_S(\bi{
q},\bi{k})|^2$ components (not plotted) is around 50\%. 
It is due to \eref{CCCampsym} applied to \eref{approx} that allows for 
such good agreement between the coherent and incoherent prescriptions.

Looking at the case $\theta_{AB}=80^\circ$ the agreement between the
CCC theory and experiment appears satisfactory. However, increasing
the difference between the two detectors by just $10^\circ$ results in 
a large rise in the experimental TDCS in the region of $20^\circ$ and
$60^\circ$ degrees. This is not reproduced by either theory, both of
which predict only a marginal increase in the TDCS. In going from
$\theta_{AB}=90^\circ$  to $\theta_{AB}=100^\circ$ both theories and
experiment predict a small increase in the TDCS, with the discrepancy
in the $20^\circ$ to $80^\circ$ angular range remaining. Increasing
$\theta_{AB}$ by  
$20^\circ$ more results in the experimental TDCS at $20^\circ$ and
$70^\circ$  to drop substantially in magnitude similarly to the CCC
theory. Curiously, if all of the
$\theta_{AB}=90^\circ,100^\circ,120^\circ,150^\circ$ 
measurements in the region of $20^\circ$ 
to $120^\circ$ degrees were reduced by a factor of 0.7 or so very good 
agreement with the CCC theory would be obtained. 

For variety we have also given results for the $\theta_A=45^\circ$
geometry. It is interesting since in the region of 
$\theta_B=-45^\circ$ the singlet component goes through a maximum
while the triplet goes to zero due to antisymmetry, resulting in a
triply peaked spin-averaged TDCS.

The so-called doubly symmetric ($E_A=E_B,\, \theta_A=-\theta_B$)
geometry provides a good overall test of how well the 
CCC formalism is working. The two terms in \eref{CCC} are identical
(TDCS has $\cos\theta$ dependence, hence independent of $\pm\theta$).
The triplet amplitude should be
identically zero at all angles due to
the Pauli principle, while the singlet amplitude should be zero at forward
and backward angles due to the electron-electron repulsion. Looking at
this case we see that the triplet thin curve is near zero at most
scattering angles, but rises at the forward angles. The coherent
combination, on the other hand, yields identically zero for the
triplet cross section as desired. To trace
the source of the problem is quite simple. By generating the TDCS
after each partial wave $J$ of total orbital angular momentum we find that 
the forward triplet TDCS grows for $J>5$. This is because exchange may 
only be treated properly between electron functions of same angular
momentum. Given that we have $l_{\rm max}=5$, for higher $L$ of
the projectile exchange cannot be fully implemented. Though presently
not practical, for computational reasons, larger $l_{\rm max}$ would be
necessary to obtain even smaller triplet TDCS.

Overall, we find the agreement with experiment in this case somewhat
disturbing. Here the excess energy is 16.4~eV. It is interesting to
compare with the 44.6~eV e-He ionization case, where the excess energy
is 20~eV \cite{Rioual98}. Generally much better agreement with
experiment is found in this case, particularly at
$\theta_{AB}=90^\circ$. Incidentally, the rescaling of the CCC theory
in the latter case was independently found to be a factor of two.

\subsection{Incident electron energy 27.2 eV}
This energy is particularly interesting due to experimental data being
available at $E_B=2$~eV \cite{BRBE96}, $E_B=4$~eV \cite{ER97} and
$E_B=E_A=6.8$~eV \cite{BBKetal91} secondary 
energies. Unfortunately the data are relative and may not be related
across the energy-sharing. This case has been recently studied by
\citeasnoun{JM98} and \citeasnoun{BBBF99}. The latter
presented the 3C, DS3C and a CCC(15,5) calculation, and suggested that
the calculations of \citeasnoun{JM98} may be much too low. Here we
present the results of a CCC(18,5) calculation. Its results are
compared to those of the CCC(15,5) calculation to test both the
convergence and the rescaling prescription.

\begin{figure}
\hskip2truecm\epsffile{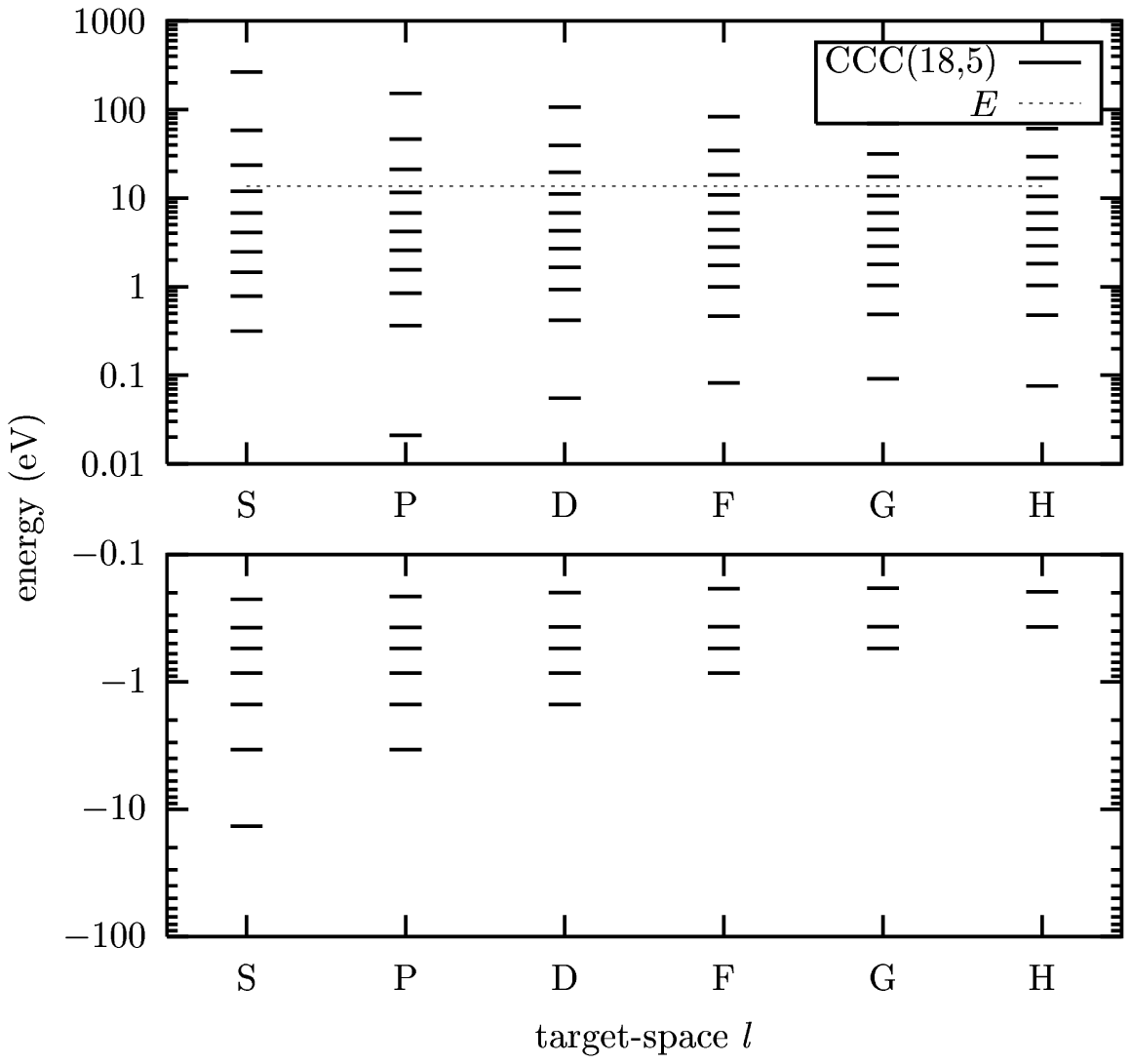}
\caption{The energy levels $\epsilon^{(N)}_{nl}$ arising in the 27.2~eV
e-H calculation using the CCC(18,5) model with
$\lambda_l\approx1.0$. The $\lambda_l$ were chosen so that for each
$l$ one energy was 6.8~eV.
\label{27.2en}
}\end{figure}
\begin{figure}
\epsffile{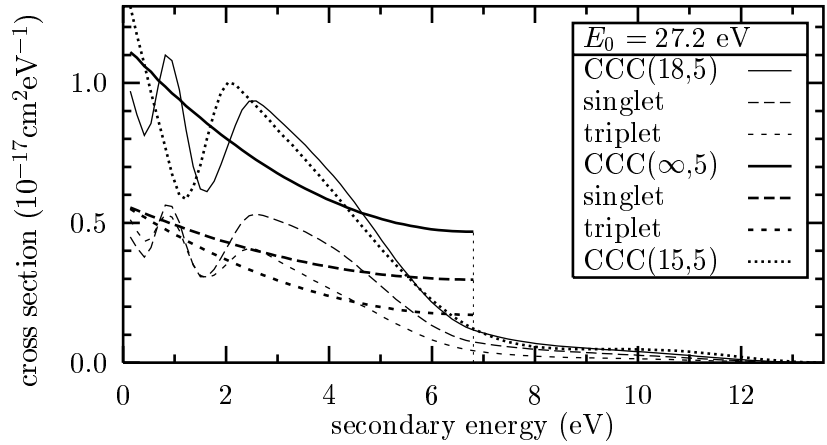}
\caption{The singly differential cross section for 27.2~eV
electron-impact ionization of the ground state of atomic hydrogen. 
The singlet and triplet (spin weights included) results are obtained
directly from the 
CCC(18,5) calculation c.f. \eref{intE}. The
CCC($\infty,5$) curve is an integral preserving estimate, see text. 
The CCC(15,5) curve is from \protect\citeasnoun{BBBF99}. The ratios
of CCC($\infty,5$) to CCC(18,5) at 2, 4 and 6.8~eV are 1.0, 0.8 and 4.
\label{27.2sdcs}
}\end{figure}
The energy levels of this calculation are given in \fref{27.2en}.
They differ substantially from those used in the CCC(15,5) calculation 
\cite{BBBF99}, and thus provide for a particularly good test of the CCC
formalism for increasing $N_0$. The $\lambda_l$ were chosen so that
one of the energies was equal to 6.8~eV for each $l$.

In \fref{27.2sdcs} the SDCS arising from the CCC(18,5)
calculation are considered. No experimental SDCS are available at this
energy, but we compare with the SDCS arising from the CCC(15,5)
calculation \cite{BBBF99}.
The discussion of the 30~eV SDCS is equally applicable here, including 
the estimation of CCC($\infty,5$). 
Some difference can be seen between the CCC(18,5) and CCC(15,5) SDCS, with the
former showing more oscillation than the latter. Yet the two SDCS are
nearly identical at $E/2$. 

\begin{figure}
\hspace{-2truecm}\epsffile{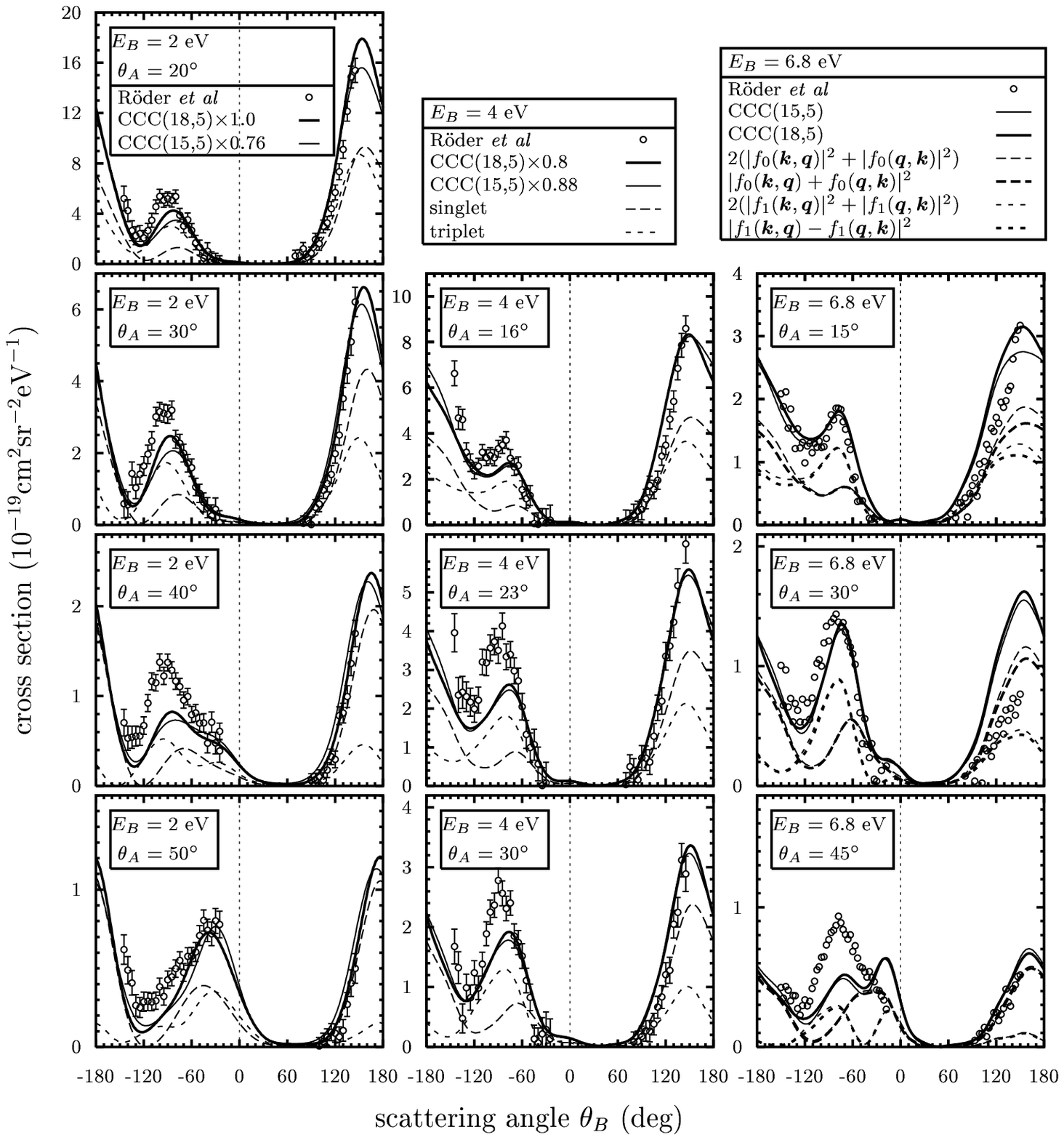}
\caption{The coplanar triply differential cross
sections for 27.2~eV 
electron-impact ionization of the ground state of atomic hydrogen. The 
internormalized relative $E_B=$2, 4, and 6.8~eV measurements are from
\protect\citeasnoun{BRBE96}, \protect\citeasnoun{ER97} and
\protect\citeasnoun{BBKetal91}, respectively. The measurements have
been normalised using a single $E_B$-dependent factor,
to the CCC(18,5) calculation. The CCC(15,5)
TDCS is from  \protect\citeasnoun{BBBF99}, however at 6.8~eV, like the
CCC(18,5) TDCS, has been obtained using the right side of \protect{approx}.
\label{27.2tdcs}
}
\end{figure}

The corresponding TDCS are presented in
\fref{27.2tdcs}. 
We see that in all cases the agreement between the two CCC
calculations is very good, confirming the claim of relatively fast
convergence in the angular distributions generally, and absolute
values, so long 
as account is taken that convergence of the raw CCC results at $E/2$
is to half the true magnitude, with subsequent rescaling. The
presented calculations are also a good check of 
the internal interpolation \cite{BF96} which is necessary in both
calculations for $E_B$=2 and 4~eV.

Though the CCC-calculated TDCS have converged (again $l_{\rm max}=5$
is sufficient) in both shape and magnitude (after rescaling),
occasional substantial 
discrepancy with experiment is disturbing. A case we would like to
single out is for $E_B=E_A, \theta_A=45^\circ$. Here, as at 30~eV, around
$\theta_B=-45^\circ$ the singlet TDCS goes through a maximum while the 
triplet TDCS goes through zero. This leads to a triply peaked
CCC-calculated TDCS, contrary to the experimental
finding. Furthermore, the DS3C calculation (see \citeasnoun{BBBF99}) is in much
better agreement with experiment than the CCC calculations. We have no 
explanation for this. Since \eref{approx} is well-satisfied the
problem is not due symmetry problems in the amplitudes.
Whereas agreement with experiment is
satisfactory at $\theta_A=15^\circ$ and  $\theta_A=30^\circ$ such
discrepancy for $\theta_A=45^\circ$ is surprising. For other cases 
the agreement with experiment is generally satisfactory.

\subsection{Incident electron energy 25 eV}
\begin{figure}[b]
\hskip2truecm\epsffile{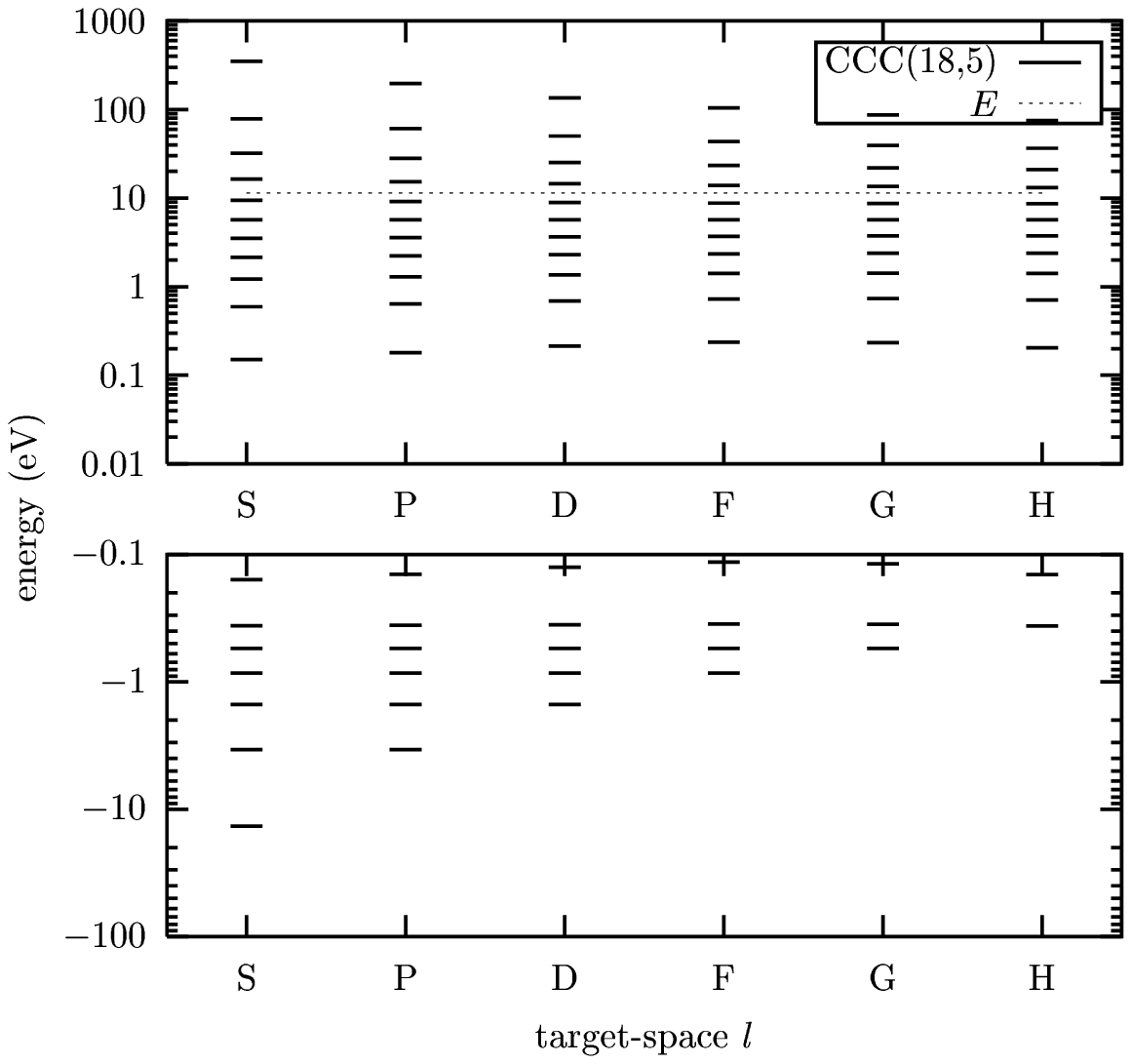}
\caption{The energy levels $\epsilon^{(N)}_{nl}$ arising in the 25~eV
e-H calculation using the CCC(18,5) model with
$\lambda_l\approx1.0$. The $\lambda_l$ were chosen so that for each
$l$ one energy was 5.7~eV.
\label{25en}
}\end{figure}

At 25~eV coplanar equal energy-sharing relative fixed $\theta_{AB}$
data are available  
\cite{Roeder96} as well as for the symmetric geometry \cite{Whelan94}.
In \fref{25en} the energy levels of the CCC(18,5) calculation are
presented. For each $l$ there is a state with energy $E/2=5.7$~eV.

\begin{figure}
\epsffile{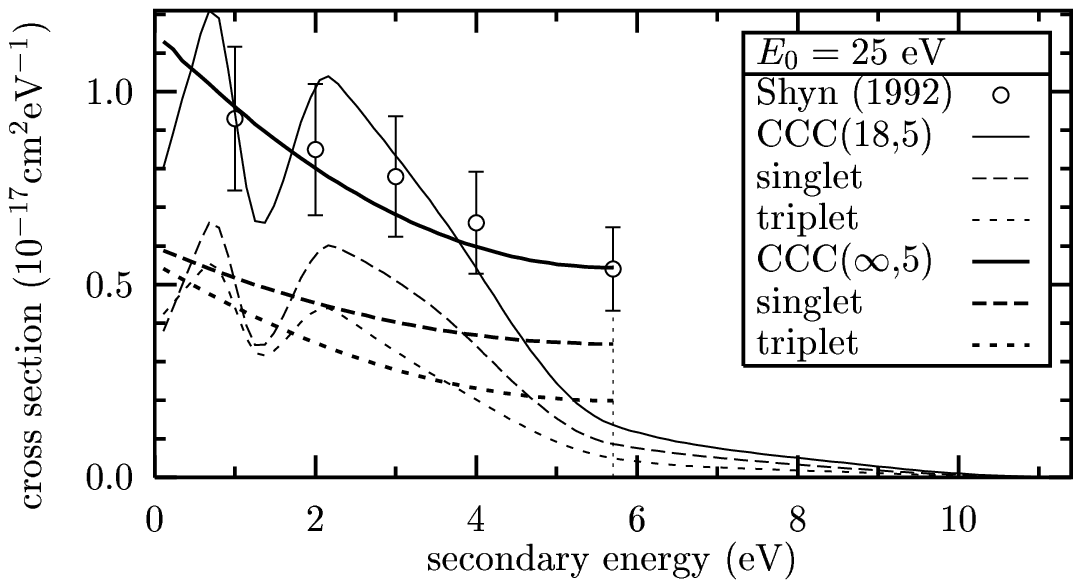}
\caption{The singly differential cross section for 25~eV
electron-impact ionization of the ground state of atomic hydrogen. 
The singlet and triplet results are obtained directly from the
CCC(18,5) calculation c.f. \eref{intE}. The
CCC($\infty,5$) curve is an integral preserving estimate, see
text. The singlet and triplet contributions include the spin weights.
\label{25sdcs}
}\end{figure}

In \fref{25sdcs} the SDCS arising from the CCC(18,5)
calculation are considered and compared with the data of
\citeasnoun{Shyn92}. 
Once again the discussion of the 30~eV SDCS is equally applicable
here. We see good agreement of the CCC($\infty,5$) estimate (see
above) with the experimental data, which at this energy has not been
rescaled as it is already in agreement with the data of
\citeasnoun{SEG87}, see \fref{ics}.

\begin{figure}
\epsffile{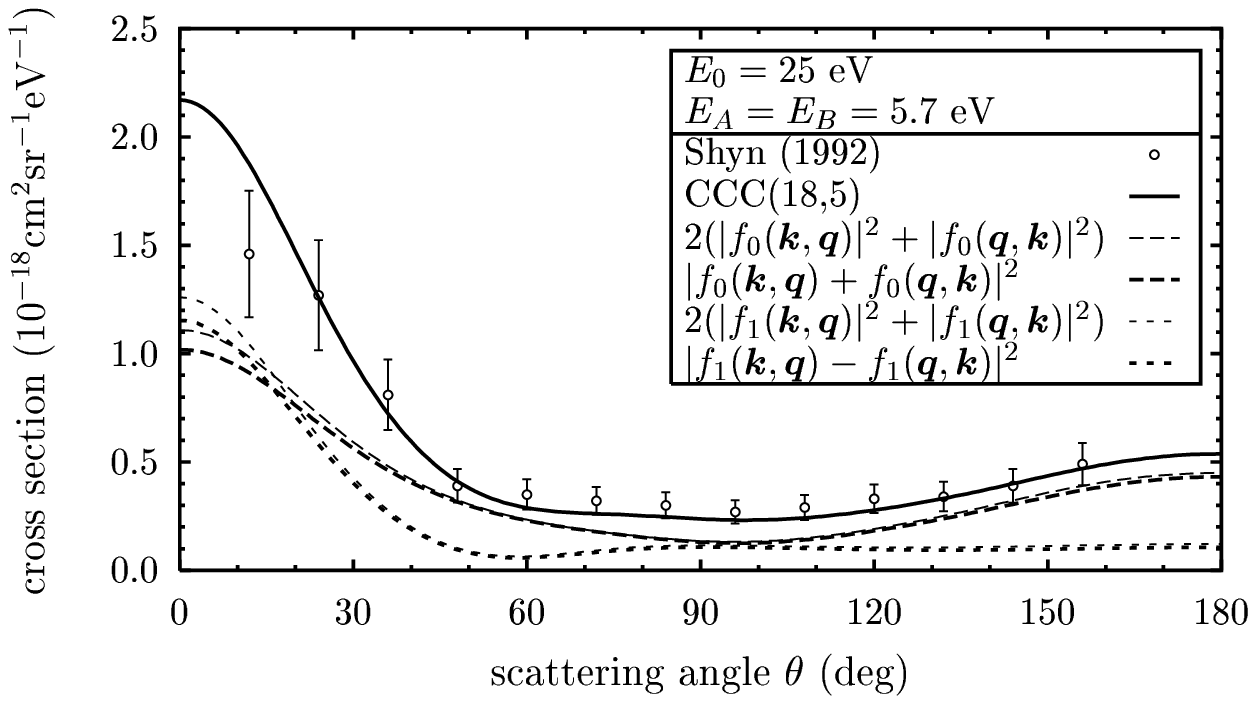}
\caption{The doubly differential cross section of the 5.7~eV
outgoing electrons for 25~eV
electron-impact ionization of the ground state of atomic hydrogen. The
singlet and triplet contributions include the spin weights, and have
been evaluated using both sides of \protect\eref{approx} prior to
integration over one of the $d\Omega$.
\label{25ddcs}
}
\end{figure}

The 25 eV DDCS are given in \fref{25ddcs} and are compared with experiment.
\begin{figure}
\hspace{-2truecm}\epsffile{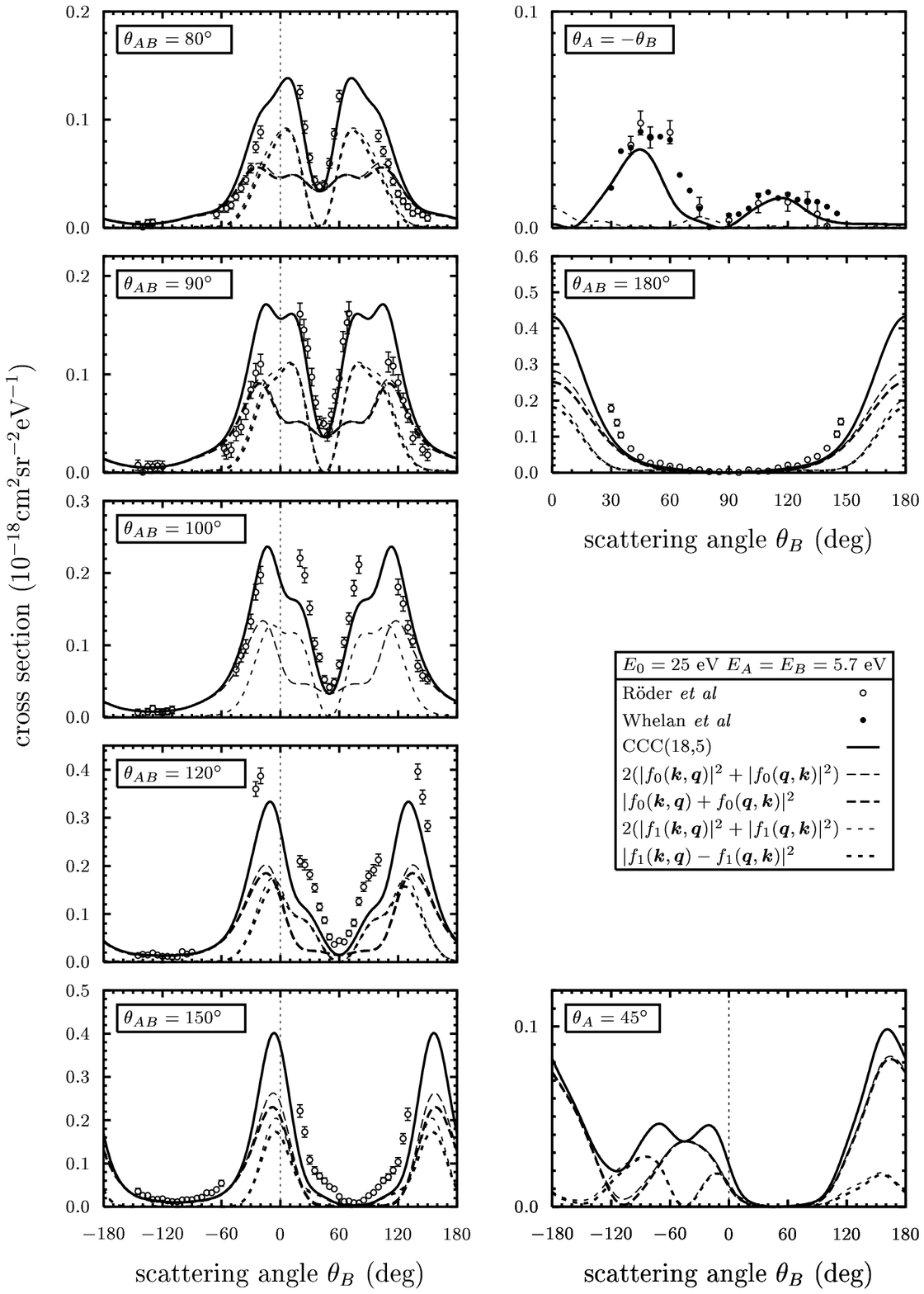}
\caption{The coplanar equal energy-sharing triply differential cross
sections  for 25~eV
electron-impact ionization of the ground state of atomic hydrogen. The 
internormalized relative $\theta_{AB}$ measurements, due to
\protect\citeasnoun{Roeder96}, have been normalised by a single factor 
to the  CCC(18,5) calculation, whose singlet and triplet (with
weights) components are evaluated using \protect\eref{approx}. The measurements
presented by 
\protect\citeasnoun{Whelan94} are internormalized with those of
\protect\citeasnoun{Roeder96}.
\label{25tdcs}
}
\end{figure}
This time we find complete agreement with experiment. Why this should
be so at this, relatively low, energy and not at higher ones is a
somewhat surprising, and may be coincidental. Once again very good
agreement between the two sides of \eref{approx} is found for both the 
singlet and triplet components.

The TDCS are presented in
\fref{25tdcs}. The coplanar relative $\theta_{AB}$ measurements of
\citeasnoun{Roeder96} have been 
scaled by a single 
factor for best overall visual fit to the theory. In order to
internormalize the coplanar 
symmetric data presented by \citeasnoun{Whelan94} we have extracted
the symmetric geometry points from the $\theta_{AB}$ measurements. 
The general agreement with experiment is not too bad. The transition
from $\theta_{AB}=80^\circ$ to $\theta_{AB}=90^\circ$ is now more
consistent than in the case of 30~eV incident energy. Interestingly,
as at 30~eV, a systematic reduction of the measurements in the
$20^\circ$-$80^\circ$ region relative to others would result in even
better agreement with experiment. The decomposition of the CCC results 
into their singlet and triplet components is helpful to check the
accuracy of the coherent versus incoherent combinations of the CCC
amplitudes, see \eref{approx}.

\subsection{Incident electron energy 20 eV}
\begin{figure}[b]
\hskip2truecm\epsffile{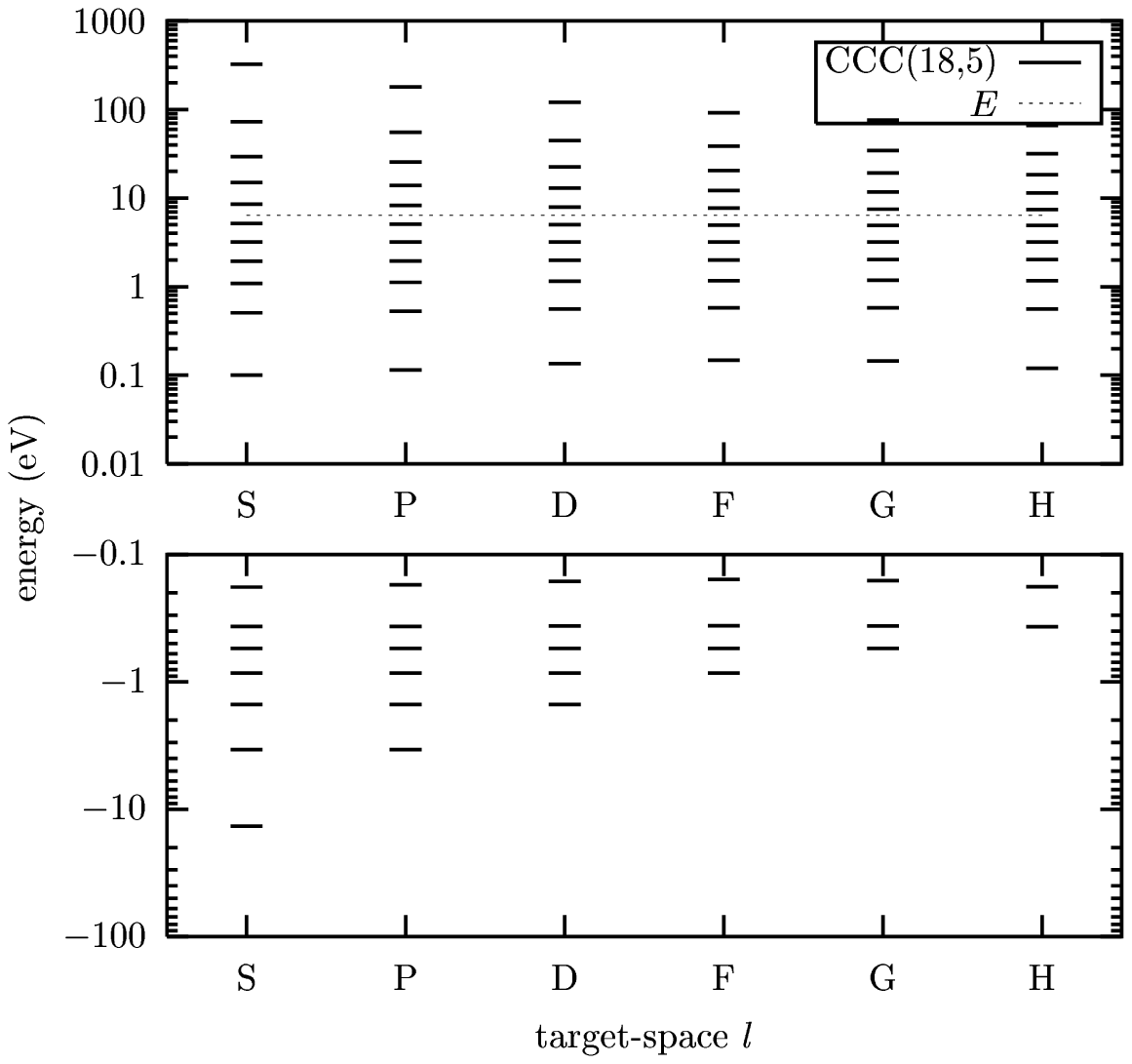}
\caption{The energy levels $\epsilon^{(N)}_{nl}$ arising in the 20~eV
e-H calculation using the CCC(18,5) model with
$\lambda_l\approx1.0$. The $\lambda_l$ were chosen so that for each
$l$ one energy was 3.2~eV.
\label{20en}
}\end{figure}
The availability of the 20~eV incident energy measurements is much the
same as for the 30 and 25~eV cases. 
Coplanar data are available for equal energy-sharing relative fixed
$\theta_{AB}$ and symmetric geometries  \cite{Roeder96,Whelan94}.

\begin{figure}
\epsffile{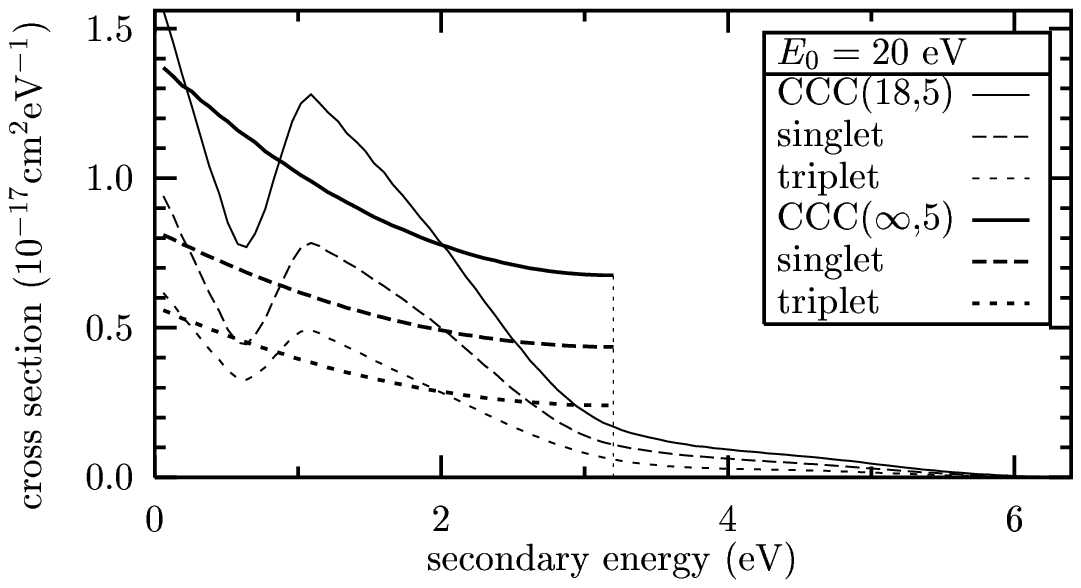}
\caption{The singly differential cross section for 20~eV
electron-impact ionization of the ground state of atomic hydrogen. 
The singlet and triplet results are obtained directly from the
CCC(18,5) calculation c.f. \eref{intE}. The
CCC($\infty,5$) curve is an integral preserving estimate, see
text.  
\label{20sdcs}
}\end{figure}
\begin{figure}
\epsffile{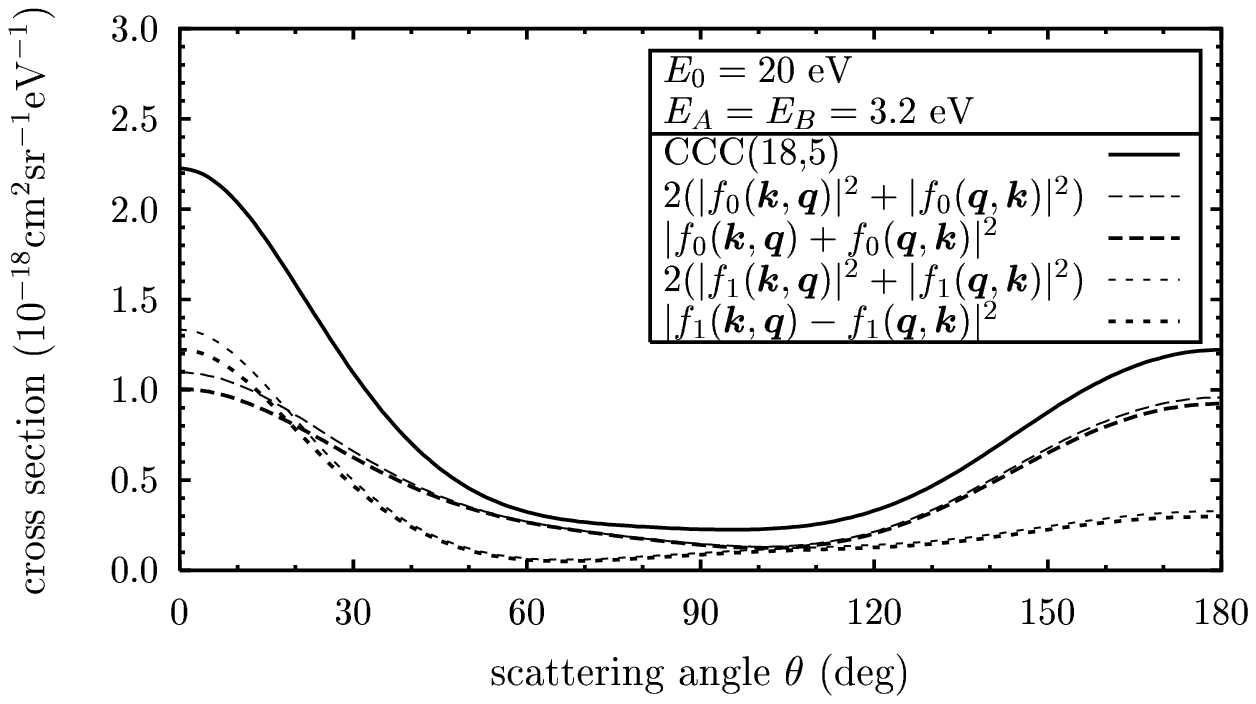}
\caption{The doubly differential cross section of the 3.2~eV
outgoing electrons for 20~eV
electron-impact ionization of the ground state of atomic hydrogen. The
singlet and triplet contributions include the spin weights, and have
been evaluated using both sides of \protect\eref{approx} prior to
integration over one of the $d\Omega$.
\label{20ddcs}
}
\end{figure}
\begin{figure}
\hspace{-2truecm}\epsffile{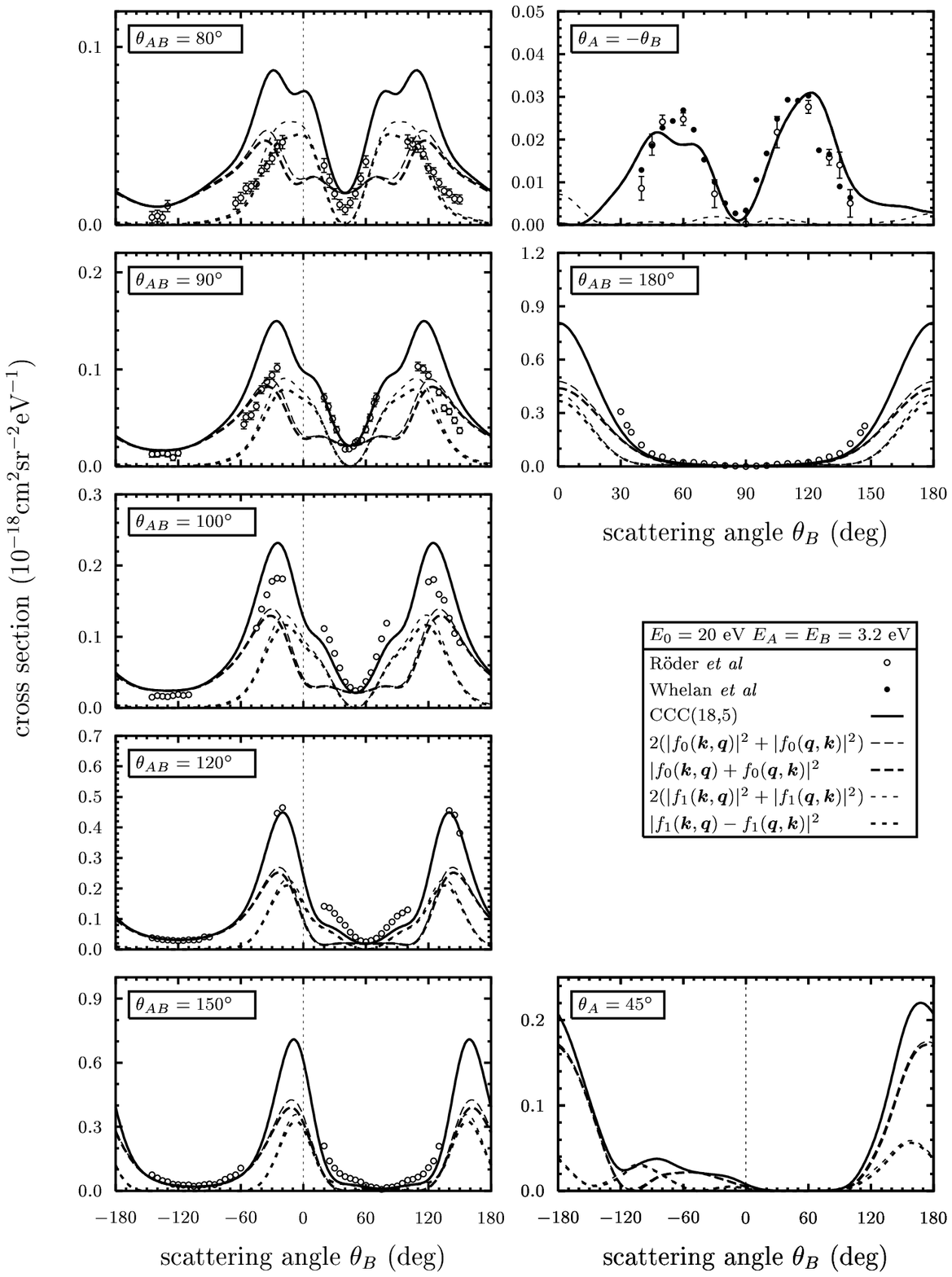}
\caption{The coplanar equal energy-sharing triply differential cross
sections  for 20~eV
electron-impact ionization of the ground state of atomic hydrogen. The 
internormalized relative $\theta_{AB}$ measurements, due to
\protect\citeasnoun{Roeder96}, have been normalised by a single factor 
to the CCC(18,5) calculation, whose singlet and triplet (with weights) 
components have been evaluated using \protect\eref{approx}. The measurements,
presented by 
\protect\citeasnoun{Whelan94}, have been internormalized with those of
\protect\citeasnoun{Roeder96}.
\label{20tdcs}
}
\end{figure}
In \fref{20en} the energy levels of the CCC(18,5) calculation are
presented, where this time there is a state of energy 3.2~eV for each $l$.
The SDCS arising from the CCC(18,5)
calculation are presented in \fref{20sdcs}.
We see that the triplet component is now systematically lower than the
singlet, with both showing similar unphysical
oscillations. The two given integral preserving quadratic estimates of
the SDCS are not used in the present calculations since data is only
available for the equal-energy-sharing kinematical region.

The 20 eV DDCS are given in \fref{20ddcs}. No experiment is yet
available, and so we present it for completeness in the hope that this 
work will generate some interest in measuring these fundamental
cross sections on a broad energy range.

The TDCS are presented in
\fref{20tdcs}. As at 30 and 25~eV the relative constant $\theta_{AB}$
measurements of \citeasnoun{Roeder96} have been 
scaled by a single 
factor for best overall visual fit to the CCC(18,5)
theory. In order to internormalize the 
symmetric data \cite{Whelan94} we have extracted
the symmetric geometry points from the $\theta_{AB}$ measurements. 
We see that for the smaller $\theta_{AB}$ there is a major
problem. Though the shape of theory and experiment is generally quite
similar there is significant discrepancy in magnitude. 
We wonder if the experimental
internormalization is at least partially responsible for the
discrepancy. For small $E_A=E_B$ and small $\theta_{AB}$ the TDCS are
particularly small, and it would be helpful to have a number of fixed
$\theta_A$ geometries measured to check the consistency of the
internormalization. Because of the substantial
discrepancies we performed many calculations which included CCC(18,4)
and CCC(20,5) models. These yield barely different results, in shape
and magnitude, to those
presented. We acknowledge certain numerical difficulties with the
presented calculations as can be observed from the non-zero triplet
TDCS for the symmetric geometry calculated using the incoherent
combination of amplitudes. However, we do not believe they are
the cause of the substantial discrepancies observed here, since
generally the agreement between the two sides of \eref{approx} is very 
good.

\subsection{Incident electron energy 17.6 eV}
We now approach the near threshold region of e-H ionization. Here
absolute TDCS are available~\cite{Retal97l}. Furthermore, the
data are very detailed in that both fixed $\theta_A$ and
$\theta_{AB}$, as well as symmetric geometries have been measured.
As before, all of the data are coplanar.

\begin{figure}
\hskip2truecm\epsffile{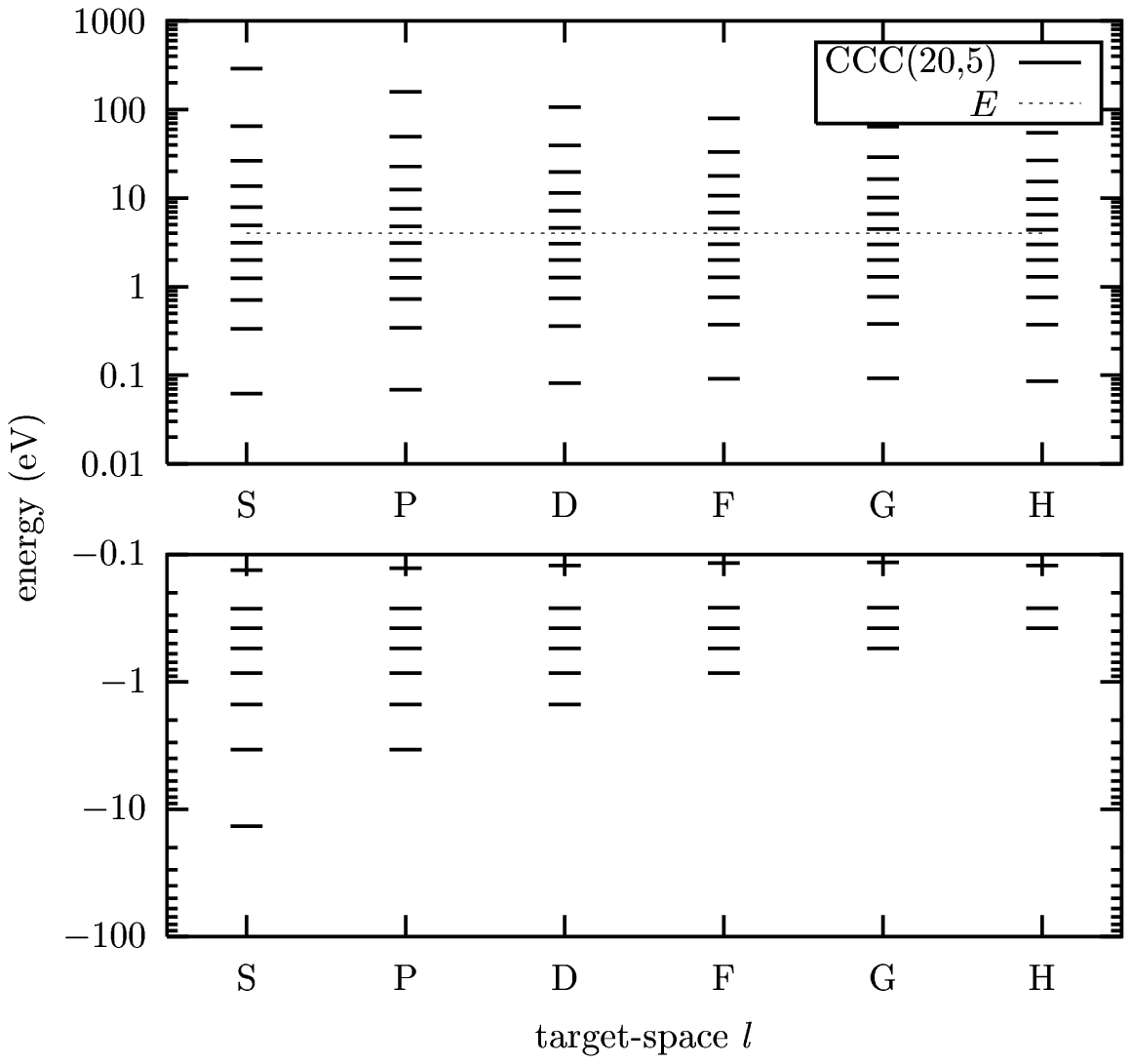}
\caption{The energy levels $\epsilon^{(N)}_{nl}$ arising in the 17.6~eV
e-H calculation using the CCC(20,5) model with
$\lambda_l\approx0.8$. The $\lambda_l$ were chosen so that for each
$l$ one energy was 2~eV.
\label{17.6en}
}\end{figure}

In \fref{17.6en} the energy levels of the CCC(20,5) calculation are
presented. 
The value of $N_0$ has been increased and the $\lambda_l$ decreased in
order to get a more accurate description of the kinematic region below 
the $E=4$~eV total energy. We also performed many smaller
calculations which show marginal difference to the largest presented.

\begin{figure}
\epsffile{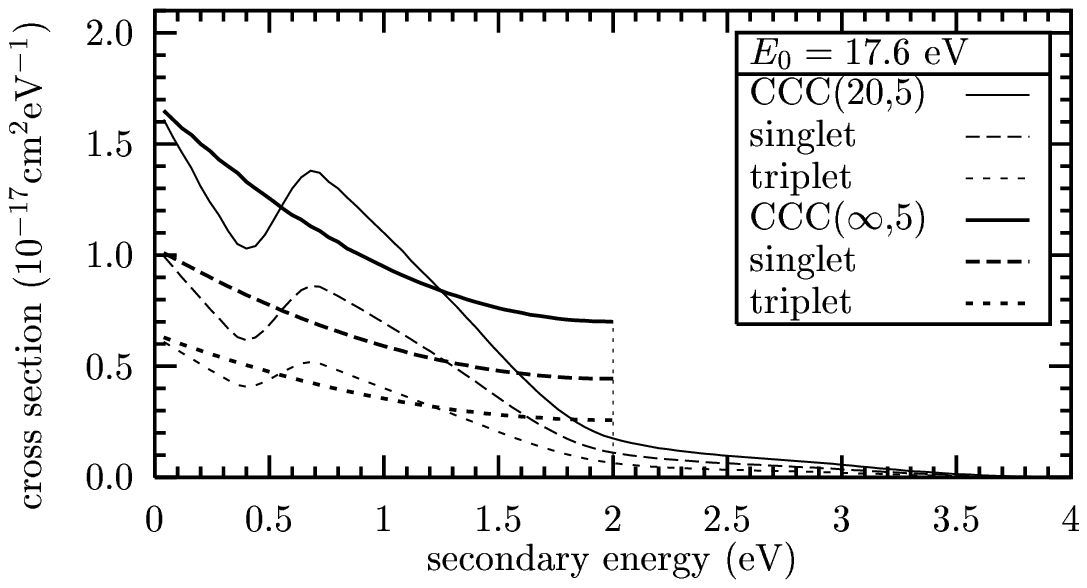}
\caption{The singly differential cross section for 17.6~eV
electron-impact ionization of the ground state of atomic hydrogen. 
The singlet and triplet results are obtained directly from the
CCC(20,5) calculation c.f. \eref{intE}. The
CCC($\infty,5$) curve is an integral preserving estimate, see
text. The
singlet and triplet contributions include the spin weights. 
\label{17.6sdcs}
}\end{figure}

In \fref{17.6sdcs} the SDCS arising from the CCC(20,5)
calculation are considered.
We see that the triplet component is now even lower than
the singlet, showing similar but less pronounced unphysical
oscillations. 

\begin{figure}
\epsffile{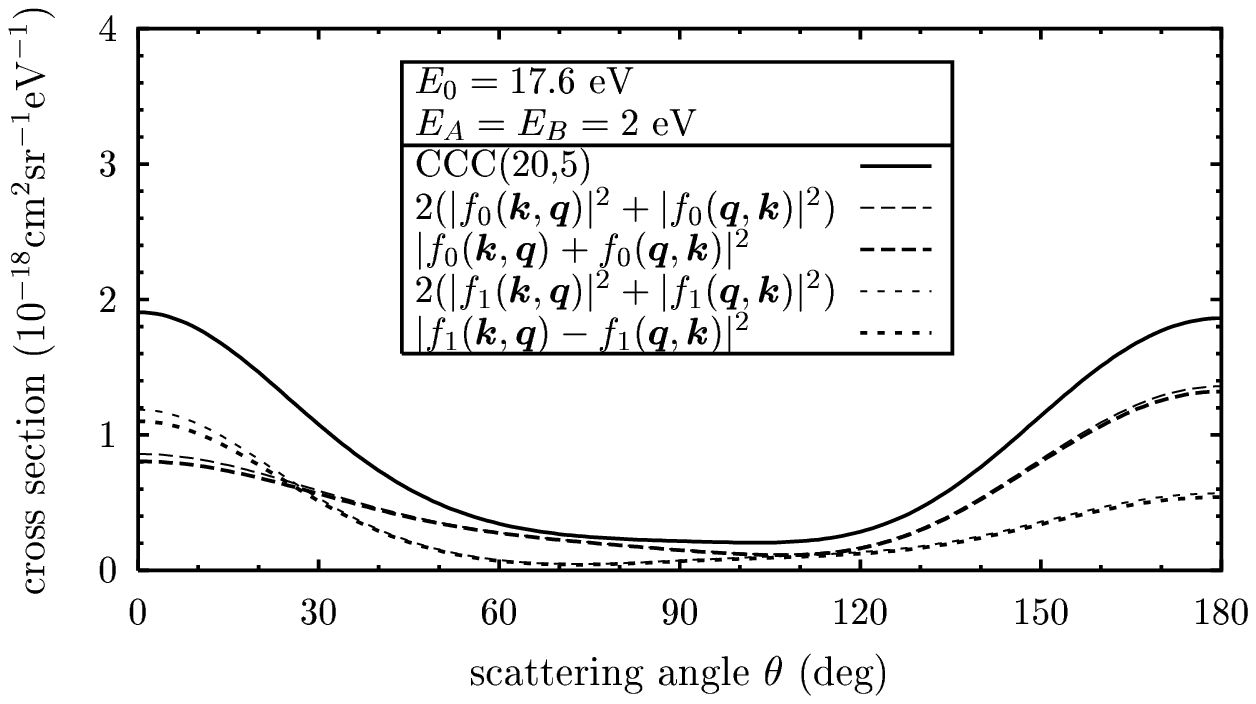}

\caption{The doubly differential cross section of the 2~eV
outgoing electrons for 17.6~eV
electron-impact ionization of the ground state of atomic hydrogen. The
singlet and triplet contributions include the spin weights, and have
been evaluated using both sides of \protect\eref{approx} prior to
integration over one of the $d\Omega$.
\label{17.6ddcs}
}
\end{figure}

For completeness the 17.6 eV DDCS are given in \fref{17.6ddcs}. 
It shows the unusual situation where forward and backward scattering
are equally dominant.
\begin{figure}
\hspace{-2truecm}\epsffile{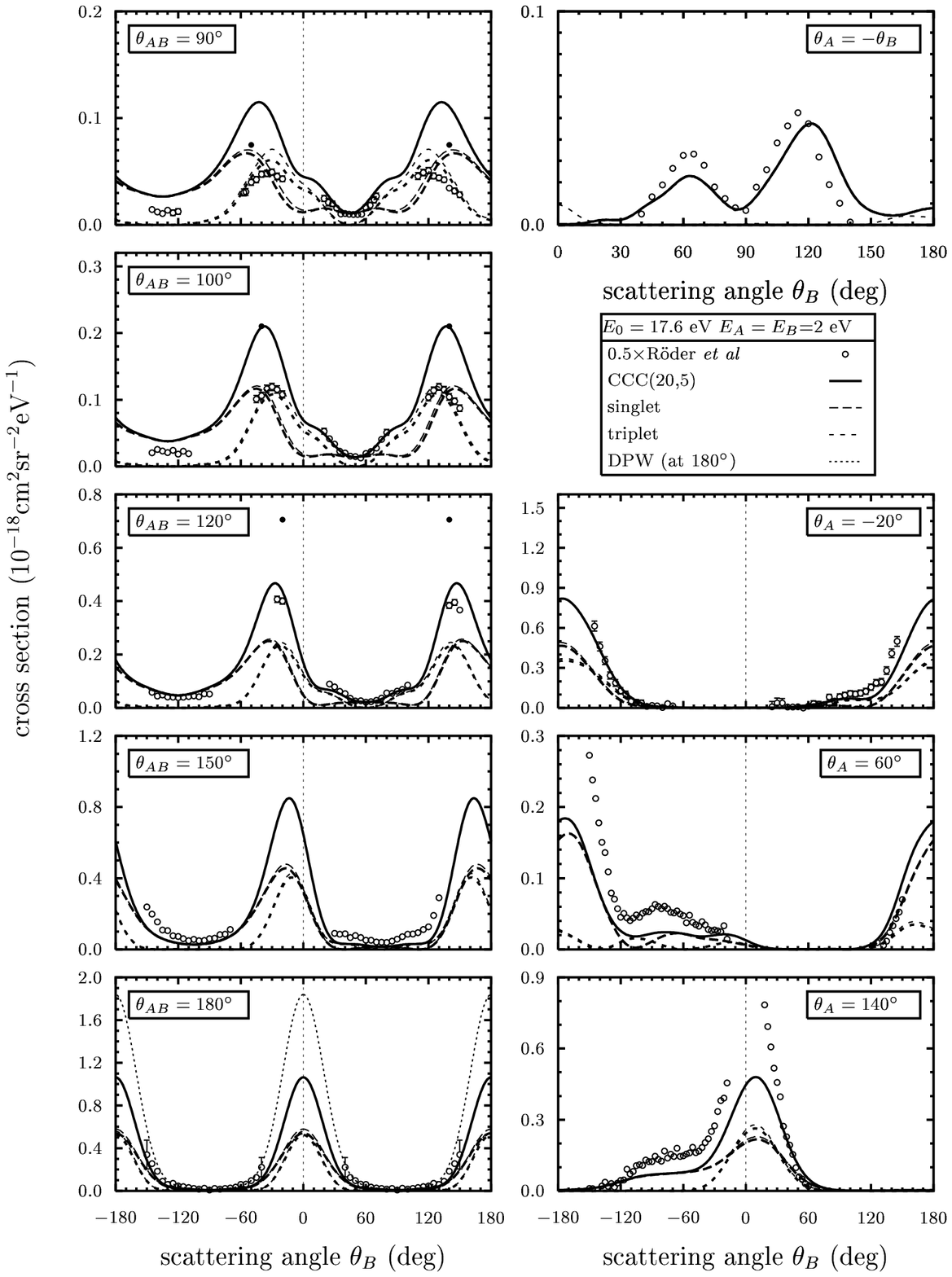}
\caption{The coplanar equal energy-sharing triply differential cross
sections  for 17.6~eV
electron-impact ionization of the ground state of atomic hydrogen. The 
open circles, denoting the absolute measurements of
\protect\citeasnoun{Retal97l} and \protect\citeasnoun{Roeder96}, have
been reduced by the 0.5 factor for best overall visual agreement
to the CCC(20,5) calculation, whose singlet and triplet (with weights) 
components are given according to \protect\eref{approx}. The solid circles
for the $\theta_{AB}=90^\circ,100^\circ,120^\circ$ are from the
$\theta_A=140^\circ$ measurement for
$\theta_B=50^\circ,40^\circ,20^\circ$, respectively. The distorted
partial-wave (DPW) calculation is due to \protect\citeasnoun{PS92} and
reported by \protect\citeasnoun{Retal97l}.
\label{17.6tdcs}
}
\end{figure}

The TDCS are presented in \fref{17.6tdcs}. In order to obtain best
visual agreement of the rescaled CCC(20,5) calculations with
experiment as a whole the measurements were 
scaled by a factor of 0.5. This is a little outside the $\pm40$\%
experimental uncertainty \cite{Retal97l}. 

As at 20~eV there are substantial discrepancies for the fixed small
$\theta_{AB}$ geometries. This time it is not just a problem of
internormalization. The discrepancy around $60^\circ$ is substantially 
smaller than at say $120^\circ$. It is particularly helpful to have so 
many different geometries measured. The symmetric geometry defines
the relationship between the singlet theoretical component and the
experiment.  The discrepancy at backward $\theta_A=-\theta_B$ angles
is responsible for the 
difference between experiment and theory in the region of $-120^\circ$ 
for the $\theta_{AB}=90^\circ,\ 100^\circ,\ 120^\circ$ geometries. The 
singlet and triplet components evaluated according to \eref{approx}
are in good agreement with each other generally. One exception is at
forward angles of the symmetric geometry where the triplet TDCS evaluated
using the left side of \eref{approx} is non-zero. The right side of
\eref{approx} yields identically zero for the triplet cross section.

We are also able to check the internal consistency of the measurements by
taking say the $\theta_A=140^\circ$ measurements and plotting them at
the appropriate points on the constant $\theta_{AB}$ plots. The solid
circles are examples of this. We see substantial inconsistency of the
measurements. The inconsistent improvement in the agreement
between theory and experiment, by simply increasing a particular
set of constant $\theta_{AB}$ measurements, implies that
internormalization is not the sole reason for the discrepancy between
theory and experiment. We hope that the presented experimental
inconsistency will lead to experimental reinvestigation of this
incident energy.

\citeasnoun{Retal97l} also presented the distorted
partial-wave (DPW) calculation of
\protect\citeasnoun{PS92}, available only for
$\theta_{AB}=180^\circ$. Comparison of the CCC results with
this calculation is also presented in \fref{17.6tdcs}. The CCC
estimate is around 1.5 times lower than the DPW calculation.

\subsection{Incident electron energy 15.6 eV}
This energy was the subject of the preliminary investigation of this
work \cite{B99jpbl}. We present these results here for completeness,
to give more information and for ready contrast to other
energies. Furthermore, the earlier results were rescaled up by a
factor of 2.7 upon the assumption of a flat true SDCS. Here we obtain
the magnitude ab initio, which indicates that the previous results
should heve been scaled up by exactly a factor of two. Hence, we
believe that the e-H SDCS is still not flat at this energy. Note that in
the case of double photoionization agreement with the SDCS($E/2$) of
\citeasnoun{PS95} implies that the CCC method is also able to predict
flat even convex SDCS.

\begin{figure}
\hskip2truecm
\epsffile{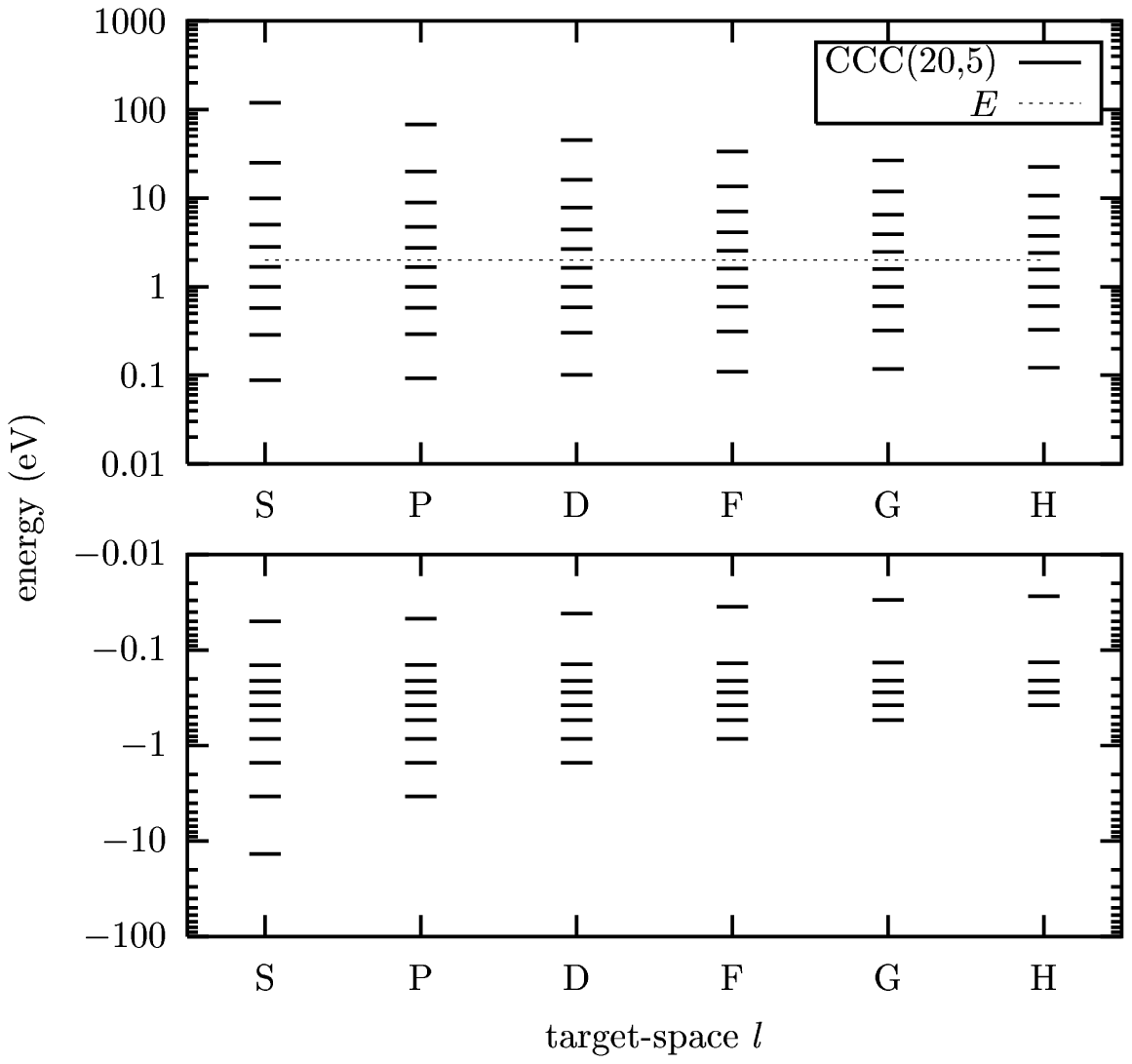}
\caption{The energy levels $\epsilon^{(N)}_{nl}$ arising in the 15.6~eV
e-H calculation ($E=2$~eV) using the CCC(20,5) model with
$\lambda_l\approx0.6$. The $\lambda_l$ were chosen so that for each
$l$ one energy was 1~eV.
\label{15.6en}
}
\end{figure}

The energies arising in the CCC(20,5) calculations are given in
\fref{15.6en}. The $\lambda_l\approx0.6$ have been reduced further
in order to have more states of energy less than the 2~eV total
energy. Though the ideal value of $\lambda_0$ for the 1S state is two,
with a basis size of 20 there is no difficulty in reproducing the 1S
state even with $\lambda_0\approx0.6$.

\begin{figure}
\hskip2truecm\epsffile{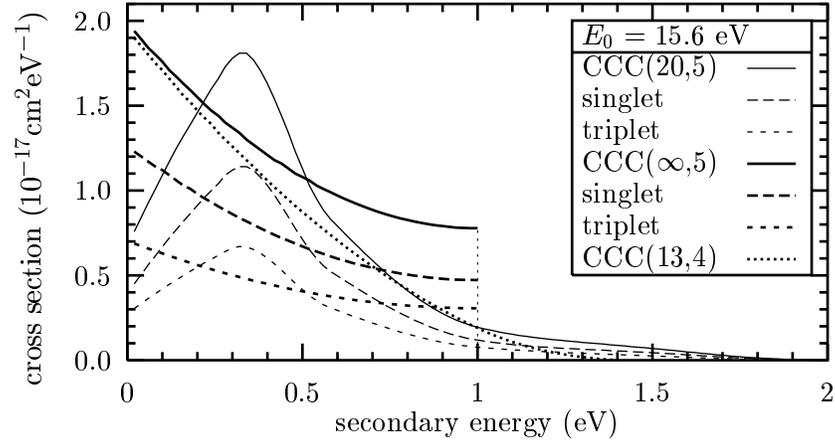}
\caption{The singly differential cross section for 15.6~eV
electron-impact ionization of the ground state of atomic hydrogen. 
The singlet and triplet results are obtained directly from the
CCC(20,5) calculation c.f. \eref{intE}. The
CCC($\infty,5$) curve is an integral preserving estimate, see text. The
singlet and triplet contributions include the spin weights. Both, the
CCC(20,5) and the CCC(13,4) are from \protect\citeasnoun{B99jpbl}.
\label{15.6sdcs}
}\end{figure}

In \fref{15.6sdcs} the SDCS arising from the CCC(20,5)
calculation are considered. Also given is the spin-averaged SDCS of
the CCC(13,4) calculation published earlier \cite{B99jpbl}. The two
agree very well at the $E/2$ point, and yield a quarter of the true
SDCS. Whereas previously we thought that this was an indication of
extremely slow convergence, now we realize that convergence has been
achieved in the CCC-calculated amplitudes, but to half the true magnitude.
The shape of the CCC-calculated SDCS has changed substantially from
the flat SDCS we supposed earlier \cite{B99jpbl}. Perhaps the work 
of \citeasnoun{BRIM99} applied to the full e-H problem will give
definitive SDCS that may be compared with the estimates given.

\begin{figure}
\epsffile{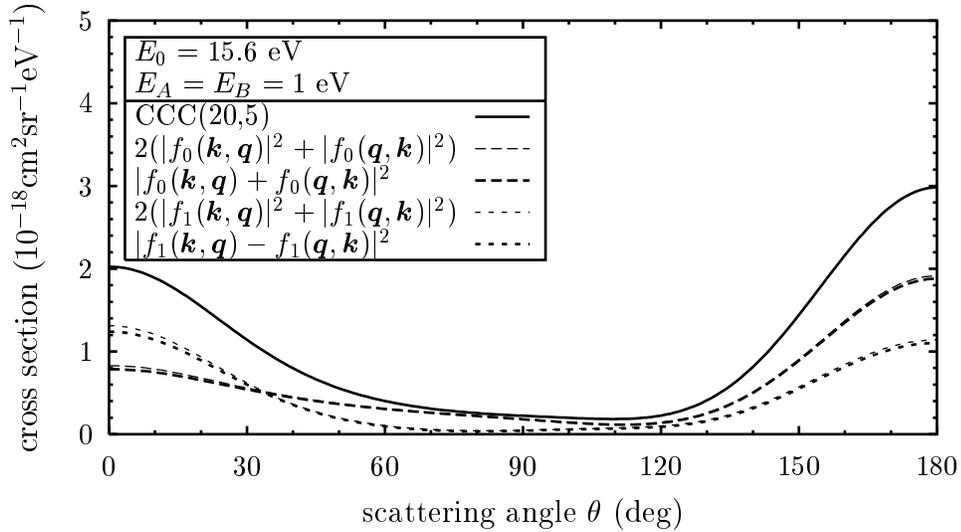}

\caption{The doubly differential cross section of the 1~eV
outgoing electrons for 15.6~eV
electron-impact ionization of the ground state of atomic hydrogen. The
singlet and triplet contributions include the spin weights, and have
been evaluated using both sides of \protect\eref{approx} prior to
integration over one of the $d\Omega$.
\label{15.6ddcs}
}
\end{figure}

The 15.6 eV DDCS are given in \fref{15.6ddcs}. Remarkably we find that
backward scattering is the most dominant.
\begin{figure}
\hspace{-2truecm}\epsffile{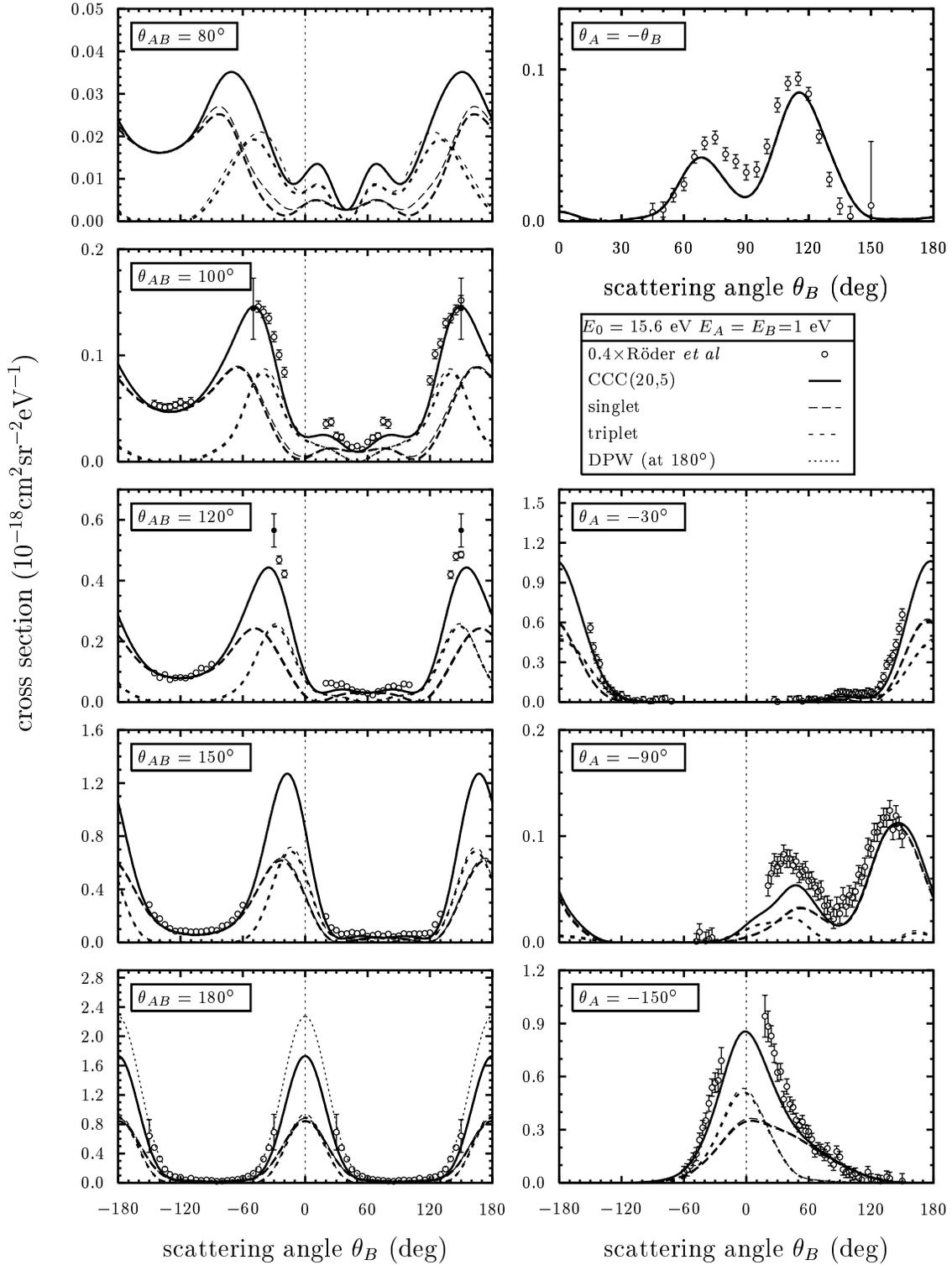}
\caption{The coplanar equal energy-sharing triply differential cross
sections  for 15.6~eV
electron-impact ionization of the ground state of atomic hydrogen. The 
absolute measurements are due to \protect\citeasnoun{Retal97l} and
\protect\citeasnoun{Roeder96}. The
solid circles for $\theta_{AB}=100^\circ,120^\circ$ geometries are
from the $\theta_A=-150^\circ$ geometry with
$\theta_B=-50^\circ,-30^\circ$, respectively.
The internormalization of the $\theta_{AB}=100^\circ$ case has been
changed from the original measurements to
the $(\theta_A,\theta_B)=(-150^\circ,-50^\circ)$ (solid circle) point,
see text. The CCC(20,5) calculation has been presented earlier
\protect\cite{B99jpbl}, but here is evaluated according to
\protect\eref{approx}.
\label{15.6tdcs}
}
\end{figure}

The TDCS are presented in \fref{15.6tdcs}. In contrast to the slightly 
higher incident energies we find excellent agreement between theory
and experiment, after the latter has been reduced by 0.4. We do note,
however, that the original internormalization of the
$\theta_{AB}=100^\circ$ measurements was not consistent with the 
$\theta_B=-50^\circ$ (solid) point of the $\theta_A=-150^\circ$
geometry. Accordingly, we imposed 
this internormalization by scaling the $\theta_{AB}=100^\circ$
measurements by a factor of 1.5 before plotting. The
$\theta_{AB}=120^\circ$ measurements are reasonably consistent with the
$\theta_B=-30^\circ$ point of the $\theta_A=-150^\circ$ geometry.

The uniform reduction of the experiment by the factor of 0.4 is
outside the stated $\pm35$~\%
uncertainty of the absolute value determination \cite{Retal97l}. The
true SDCS would have to be highly convex in order for the experimental 
absolute values to be correct. Recall that the CCC-calculated and
estimated SDCS correctly yield the spin-dependent total ionization
cross sections at this energy (see \fref{ics}).

\section{Conclusions}
We have performed an extensive and systematic study of e-H ionization
from 250~eV to 
15.6~eV incident energy. We showed how the close-coupling approach to
ionization converges to the Born approximation at high energies. While 
we believe it is common knowledge that exchange effects disappear at
high energies, \citeasnoun{BC99} argue that the treatment of exchange in
our formalism should lead to amplitudes that satisfy the
symmetrization postulate and hence yield a symmetric SDCS. Their
argument is independent of energy, and it is our
view that this claim is incorrect. Instead, we still suspect to be
true the step function hypothesis \cite{B97l}, which states that with
increasing $N$ the CCC-calculated amplitudes should 
converge to zero on the secondary energy range of $[E/2,E]$, for all
total energies $E$. The presented results are consistent with this
idea, and the unphysical oscillations in the SDCS for small $E$ being due 
to the inability of a finite expansion being able to describe a step
function of substantial step size. Thus, for any finite $N$ the
CCC-calculated ionization scattering amplitudes will generally not satisfy the
symmetrization postulate \eref{symcon}. 

The analysis of \citeasnoun{S99l} shows that at $E/2$ the
CCC-calculated amplitudes should be combined coherently. This is
consistent with the step-function hypothesis with the $E/2$ amplitudes
converging to half the step size, just like in Fourier
expansions. Accordingly, the unitarity preserving incoherent
prescription given by \citeasnoun{BF96} needs to be multiplied by two, 
but only at $E/2$. Subsequently, the two combinations of amplitudes
yield near identical results for all considered cases. This is due to
the fact that the CCC amplitudes at $E/2$ satisfy the symmetrization
postulate, at least approximately. The effect of any deviation from this on the
TDCS is particularly small, see discussion following
\eref{CCCampsym}. This reconciles the coherent versus incoherent
combinations of the total-spin-dependent CCC amplitudes as both being
effectively multiplications by two. Recall that the CCC amplitude is
already a coherent combination of its direct and exchange amplitudes
depending on the total spin.

The above discussion is only applicable to the equal-energy-sharing
kinematical region, where now we can claim to obtain fully ab initio
results convergent, using realistic calculations, in both shape and
magnitude. The situation for the asymmetric kinematical region is much 
less satisfactory. We are still unable to obtain convergence
generally at low-enough total energies $E$. The analysis of
\citeasnoun{S99l} is formally only appropriate at $E/2$, unless the CCC 
amplitudes in the region $[E/2,E]$ are identically zero. In other
words, if the step-function hypothesis is true then his work implies
that the CCC amplitudes in the region $[0,E/2]$ will be unambiguously
defined. In 
practice, when comparing with experiment the step-function idea is
well satisfied as we find that 
$|f^{(N)}_S(\bi{k},\bi{q})| \gg |f^{(N)}_S(\bi{q},\bi{k})|$ for
$q<k$. Hence, a coherent or an incoherent combination makes no
discernible difference from just using the amplitude $f^{(N)}_S(\bi{
k},\bi{q})$.

Comparison with experiment is somewhat mixed. We find it particularly
disturbing that the fundamental e-H DDCS have not been accurately
determined experimentally. We make this claim by reference to the
inconsistency between the data of \citeasnoun{Shyn92} and
\citeasnoun{SEG87}. Consistency between the present results and those
of \citeasnoun{BK93} further supports this claim.
In our view it is more important to obtain
accurate DDCS, preferably absolute, than performing more complicated
TDCS experiments. In support of this we have given an extensive
spin-resolved set of DDCS for future comparison.

Turning our attention to the TDCS we find the agreement with
experiment somewhat inconsistent. At high energies the agreement is
generally satisfactory. This varies, sometimes quite substantially, as 
the incident energy is reduced. We believe that the CCC results presented
accurately reflect the close-coupling approach to ionization in that
further even larger calculations, when computer resources permit, will
not yield substantially different results. There is some uncertainty
associated with the semi-empirical rescaling of the cross sections for 
asymmetric energy-sharing kinematics. However, given the
nature of some of the discrepancies, at this point, this is the least
of our concerns. The fundamental question we have is whether or not
the close-coupling approach to ionization, as we have defined it,
converges to the true TDCS. The result of the present study suggests
that this is still an open question. Further measurements,
particularly in order to eliminate the presented experimental
inconsistencies, would be very welcome, and help answer this
question. 

While it is clear that the close-coupling formalism is
unable to yield accurate SDCS for small enough $E$ this does not
necessarily affect the angular profiles of the TDCS as discussed
earlier \cite{B99jpbl}. The equivalent-quadrature idea in application
to the systematic generation of the square-integrable states helps to
ensure rapid convergence in the angular profiles. This may be readily
checked numerically, as we have here in \fref{27.2tdcs} for 27.2~eV
and did earlier at 15.6~eV \cite{B99jpbl}. The utility of the
rescaling prescription depends on the accuracy of the estimate of the
true spin-resolved SDCS. Should this become known, as appears likely
\cite{BRIM99}, then more accurate rescaling may be performed than what
was presented here. This, however, is only applicable to the
asymmetric energy-sharing kinematics. At equal energy-sharing we are
no-longer free to rescale our results as previously thought.

The great strength of the close-coupling approach to ionization is
that it unifies the treatment of both the discrete and continuum parts 
of the atomic spectrum. We have already established the importance of
treating the target continuum in application to discrete excitation
processes \cite{B94}. Similarly, we suspect that discrete excitation
processes need to be treated in order to assure accuracy of ionization 
calculations at all energies. We certainly hope that the present work
will stimulate further e-H ionization measurements and calculations,
and therefore test 
the ability of the present implementation of the CCC theory to be predictive.

\ack
The author is indebted to Andris Stelbovics for communicating his
findings prior to publication, and the many following discussions. We
are also grateful to Dmitry Fursa for stimulating discussions and
technical support.
We thank Colm Whelan and Jens Rasch  for providing their 30~eV data in
electronic form. 
Support of the Australian Research Council
and the Flinders University of South Australia is acknowledged.
We are also indebted to the South Australian
Centre for High Performance Computing and Communications.  

\newpage
\section*{References}

\end{document}